\documentclass[useAMS,usenatbib]{mn2e}

\usepackage{graphicx}
\usepackage{rotating}
\usepackage{amssymb}

\begin{document}
%
%
\title[A new analysis of the WASP-3b system.]{A new analysis of the WASP-3 system: no evidence for an additional companion}
\author[M. Montalto et al.]{
\parbox[t]{\textwidth}{\center M. Montalto,$^{1}$\thanks{E-mail:Marco.Montalto@astro.up.pt} 
J. Gregorio,$^{2}$ G. Bou\'e,$^{1}$ A. Mortier,$^{1}$ I. Boisse,$^{1}$ M. Oshagh,$^{1,3}$ \\ M. Maturi,$^{4}$ P. Figueira,$^{1}$
S. Sousa$^{1}$ and N. C. Santos$^{1,3}$}\vspace{0.4cm}\\\\
$^{1}$ Centro de Astrofisica da Universidade do Porto (CAUP), Rua das Estrelas, 4150-762, Porto, Portugal\\
$^{2}$ Atalaia group, Crow Observatory-Portalegre, 7300 Portalegre, Portugal\\
$^{3}$ Departamento de Fisica e Astronomia, Faculdade de Ci\^encias, Universidade do Porto, 4169-007, Porto, Portugal\\
$^{4}$ Zentrum fuer Astronomie, ITA, Universitaet Heidelberg, Albert-Ueberle-Str. 2, D-69120, Heidelberg, Germany\\
\\
}

\date{}

\pagerange{\pageref{firstpage}--\pageref{lastpage}} \pubyear{2011}

\maketitle

\label{firstpage}

\begin{abstract}
In this  work we  investigate the problem  concerning the  presence of
additional bodies gravitationally bounded  with the WASP-3 system.  We
present eight  new transits of  this planet gathered between  May 2009
and  September  2011  by  using  the  30-cm  Telescope  at  the  Crow
Observatory-Portalegre,  and analyse  all the  photometric  and radial
velocity  data  published so  far.   We  did  not observe  significant
periodicities in  the Fourier spectrum  of the observed minus 
calculated (O-C) transit  timing and
radial  velocity  diagrams (the  highest  peak  having  false  alarm
probabilities equal to 56 per cent  and 31 per cent 
respectively) or long term
trends.  Combining  all the  available information, we  conclude that
the radial  velocity and transit  timing techniques exclude  at 99
per cent
confidence limit  any perturber more  massive than $\rm M  \gtrsim 100
\,\rm M_{earth}$ with periods up to  ten times the period of the inner
planet. We  also investigate  the possible presence  of an  exomoon on
this system  and determined  that considering the  scatter of  the O-C
transit  timing  residuals a  coplanar  exomoon  would likely  produce
detectable transits.  This hypothesis  is however apparently ruled out
by observations conducted by other  researchers.  In case the orbit of
the  moon is  not  coplanar the  accuracy  of our  transit timing  and
transit  duration  measurements  prevents any  significant  statement.
Interestingly, on the basis of  our reanalysis of SOPHIE data we noted
that  WASP-3   passed  from  a  less   active  ($\log{\rm  R}^{'}_{\rm
  hk}=-4.95$) to a more active ($\log{\rm R}^{'}_{\rm hk}=-4.8$) state
during the 3 yr monitoring period spanned by the observations. Despite
no  clear spot  crossing  has  been reported  for  this system,   this
analysis claims for a more  intensive monitoring of the activity level
of  this star in  order to  understand its  impact on  photometric and
radial velocity measurements.
\end{abstract}


\section{Introduction}
\label{s:introduction}

The field  of exoplanets  is blessed in  these years by  an impressive
flow of new exciting discoveries.  As the sample of known exoplanetary
systems increases  several interesting characteristics  become evident
posing  new challanging issues  for theories  of planet  formation and
evolution.  Given  their short periods  and their large masses  the so
called  Hot-Jupiters   were  the  first  class   of  exoplanets  being
discovered around solar type stars (Mayor \& Queloz 1995). Those among
them  later  found  to transit  in  front  of  the  disk of  the  star
(Charbonneau et al.  2000) also gained a special importance given that
they allow us to acquire physical informations like the radius and the
density of  the planet which would remain  otherwise inaccessible.  At
the time  of writing  this paper, there  were 187 known  and confirmed
transiting planets,\footnote{http://exoplanets.org/table/  on March 9,
  2012} and 87 of  them have a period smaller than 10  days and a mass
larger than 0.7 $\rm M_{\rm jup}$.

Several  hypothesis were  endeavored to  explain the  origin  of these
objects   involving  scenarios  where   these  giant   planets,  while
originally  forming in remote  regions of  the planetary  system, were
then moved  to their  actual position by  means of  different possible
mechanisms which  can be essentially  grouped in three  broad classes:
(i)   planet-protoplanetary  disk   interactions  leading   to  inward
migration of the  giant planet (Goldreich \& Tremaine  1980; Nelson et
al.  2000);  (ii) planet-planet scattering  in multi-planetary systems
(Weidenschilling \&  Marzari 1996; Chatterjee et  al.  2008; Juri\'{c}
\& Tremaine 2008); (iii) Kozai-induced migration in inclined planetary
or binary stellar systems (Kozai  1962; Wu \& Murray 2003; Fabrycky \&
Tremaine 2007).  While the first  class of mechanisms would produce in
principle a smooth migration leading to circularized orbits preserving
the  original alignment  between the  spin axis  of the  star  and the
orbital angular momentum  axis of the planet, the  other two processes
may result in final eccentric orbits and largely misalinged spin-orbit
angles.  As  it was  evidenced by exploiting  the Rossiter-McLaughlin
effect (Rossiter  1924; McLaughlin  1924, hereafter RM  effect), most
transiting exoplanets have spin-orbit angles perfectly consistent with
zero, but  some of them  present surprisingly large  misaligned angles
(Triaud  et al.   2010).  These  differences highlight  the  fact that
probably  all  these processes  are  playing  a  role in  shaping  the
structure of these systems (Nagasawa et al.~2008).

To  further  clarify  the   relative  importance  of  these  different
theoretical  scenarios  and  understand  under  which  situations  one
mechanism may  prevail over the others, some  additional and important
related  questions  need to  be  carefully  examinated.   One of  them
concerns the need to understand if these objects are actually isolated
or  if other planets  or even  stellar companions  are gravitationally
bounded with the system.  Despite  the importance of this topic in the
framework of  our understanding of Hot-Jupiter  planets, our knowledge
is still far from being complete.

In this  paper, while attempting to  shed new light on  this problem, we
considered the case of  the transiting Hot-Jupiter WASP-3b, collecting
all the photometric  and radial velocity data acquired  so far as well
as presenting our new photometric measurements.  We used this database
to investigate the presence of an additional companion in this system.

WASP-3b    is    a    Hot-Jupiter    planet   with    a    mass    of
$(2.00\,\pm\,0.09)\,\rm M_{jup}$ revolving around a main sequence star
of spectral type F7-8V with a period of $\sim1.8$ days.  Its discovery
was announced in 2008 by the WASP Consortium (Pollacco et al.~2008) as
a  result of  a photometric  campaign conducted  with  the robotically
controlled WASP-North  Observatory located in La  Palma and subsequent
radial velocity follow-up obtained  with the SOPHIE spectrograph at the
Observatory de  Haute-Provence.  The  first photometry of  WASP-3b was
presented in the discovery paper  of Pollacco et al.~(2008) which used
SuperWASP-N together  with IAC80cm and  Keele 80cm telescopes  data to
refine  the  properties  of  the transiting  object.   Two  additional
transits of  WASP-3b were  observed by Gibson  et al.~(2008)  with the
RISE instrument  mounted on the fully robotic  2m Liverpool Telescope.
Tripathi et  al.~(2010) observed six  transits of WASP-3b at  the 1.2m
FLOW telescope and at the 2.2m University of Hawaii Telescope. Joining
their  results with  those of  Pollacco  et al.~(2008)  and Gibson  et
al.~(2008)  they   concluded  that  a  linear  fit   to  the  observed
ephemerides  was not  satisfactory, and  that either  the  errors were
underestimated or  there was a genuine  period variation.  Maciejewski
et  al.~(2010) presented  six new  transits gathered  at two  1m class
Telescopes (Jena and  Rozhen) pointing out that a  periodic signal was
present  in  the observed  minus  calculated  (hereafter O-C)  transit
timing  diagram, and  that an  outer perturbing  planet in  the system
could have  best explained the observations. Later  on Christiansen et
al.~(2011) discussed eight new transits of WASP-3b observed during the
NASA EPOXI Mission of Opportunity. Despite the high precision of their
transit timing measurements, these data have never been used so far to
analyze    transit   timing    variations   of    WASP-3b.    Recently
Littlefield~(2011)  reported five  additional transit  measurements of
WASP-3b  which  were  observed  with  the  11-inch  Schmidt-Cassegrain
telescope  at Jordan  Hall on  the  University of  Notre Dame  Campus.
Despite the larger uncertainties  with respect to previous studies the
analysis of Littlefield apparently  provided an initial modest support
to the hypothesis of Maciejewski et al.~(2010).  Very recently Sada et
al.~(2012) obtained three  additional lightcurves of WASP-3b observing
with the KPNO visitor center 0.5m Telescope and one with the 2.1m KPNO
Telescope.

Here we present  a study of eight new  homogeneously observed transits
of   WASP-3b.     This   paper   is   structured    as   follows:   in
Sect.~\ref{s:obs},  we   present  our  observations   of  WASP-3b;  in
Sect.~\ref{s:stellar_parameters}  we derive  the stellar  paramters of
the  host  star;  in  Sect.~\ref{s:data_reduction},  we  describe  the
reduction process; In  Sect.~\ref{s:transit_analysis}, we describe our
analysis of the photometric  data; in Sect.~\ref{s:RV}, we present the
radial velocity  data.  In Sect.~\ref{s:analysis_RV},  we describe our
analysis of the radial velocity data.  In Sect.~\ref{s:oc}, we discuss
the  O-C trasit timing  diagram while  in Sect.~\ref{s:RVres}  the O-C
radial  velocity diagram.  In  Sect.\ref{s:discussion} we  discuss our
results.   Finally  in  Sect.~\ref{s:conclusions},  we  summarize  our
results and conclude.

\begin{table}
\caption{
Informations on our observing runs.
\label{tab:obs}
}
\begin{center}
\begin{tabular}{c c c c c}
\hline
Date & Epoch & Texp & Airmass range & N.images \\
\hline
15/05/2009 & 196 &  90 & 2.016-1.002 & 178 \\
13/04/2011 & 574 &  90 & 1.873-1.004 & 104 \\
26/04/2011 & 581 &  90 & 2.055-1.005 & 123 \\
02/06/2011 & 601 & 150 & 1.532-1.024 &  96 \\ 
20/07/2011 & 627 & 150 & 1.029-2.019 &  98 \\
13/08/2011 & 640 & 150 & 1.032-1.635 &  77 \\
26/08/2011 & 647 & 150 & 1.002-1.644 &  77 \\
08/09/2011 & 654 & 150 & 1.002-1.333 &  61 \\
\hline
\end{tabular}
\end{center}
\end{table}

\begin{table}
\caption{
Technical specifications of the acquisition camera.
}
\label{tab:ccd}
\resizebox{8.6cm}{!}{
\begin{tabular}{ c c }
\hline
Camera Sbig ST8XME specifications & \\
\hline
CCD           	        &  Kodak KAF-1603ME +  TI TC-237 \\
Pixel Array           	&  1530 $\times$ 1020 pix \\
CCD Size           	&  13.8 $\times$ 9.2 mm \\
Total Pixels           	&  1.6 million \\
Pixel Size           	&  9$\mu$m $\times$ 9$\mu$m \\
Full Well Capacity      &  $\sim$100,000 $^{-}e$ \\
Dark Current           	&  1 $^{-}e$/pix/sec at 0 $^{\circ}$C. \\
Readout Specifications &  \\
Shutter			&  Electromechanical \\
Exposure           	&  0.12 to 3600 sec \\
resolution              &  10 msec \\
A/D Converter	        &  16 bits \\
A/D Gain           	&  2.17 $^{-}e$/ADU \\
Read Noise           	&  15 $^{-}e$ RMS \\
Binning Modes           &  1 $\times$ 1, 2 $\times$ 2, 3 $\times$ 3 \\
Full Frame Download     &  $\sim$4 sec \\
\hline
\end{tabular}
}
\end{table}

\section{Observations}
\label{s:obs}

The    data   described    here    were   acquired    at   the    Crow
Observatory-Portalegre  in  Portugal.   Eight  different  transits  of
WASP-3b  were  observed  as  documented in  Table~\ref{tab:obs}.   The
telescope is a 30 cm aperture  Meade LX200 F10, reduced at F5.56 (1668
mm)   focal    lenght   yielding   a   total   field    of   view   of
$\sim$28$^{'}$x19$^{'}$.  The images were  acquired with a Sbig ST8XME
camera    which   technical    characteristics    are   reported    in
Table~$\ref{tab:ccd}$.   The   pixel  scale  is   1.1$^{''}$/pix.  The
exposure time was  fixed either to 90 sec or to  150 sec, the overhead
was 4 sec  due to the full frame download time.   A total number equal
to 814  images were acquired  and analyzed.  All  of them were  in the
$I$-band filter.

\section{Data reduction}
\label{s:data_reduction}

Bias  subtraction  and  flat  fielding  were performed  with  our  own
software in  a standard  manner. We construct  a master dark  image to
identify  defective  pixels in  the  image  and  applied a  bad  pixel
correction algorithm  which interpolated the values of  the bad pixels
with  those of  the surrounding  pixels.  Then  we used  {\sc daophot}
(Stetson 1987) to derive initial aperture photometry and calculate the
point spread function  (PSF) of our images. {\sc  allstar} was used to
refine magnitude  estimates and centroid positions.   We then selected
our  best seeing  image  as astrometric  reference frame.   Coordinate
transformations  among all  the other  frames and  the  reference were
calculated  using {\sc  daomatch} and  {\sc daomaster}.   We  took the
first ten best seeing images  to construct a master high S/N reference
frame with {\sc  montage2}, and a master list  of objects.  After that
centroid  positions and  magnitudes  were further  refined using  {\sc
  allframe} (Stetson  1994).  Finally we  rederive aperture photometry
for each source after subtracting the  PSF of all the other objects in
our images.   After some  experiments we decided  to set  the aperture
radius  for each  frame to  2.3 times  the value  of the  FWHM  of the
corresponding PSF.

\subsection{Corrected lightcurves}
\label{s:corrected}

We  then  constructed  the  flux  ratios  between  our  target  source
(WASP-3b) and several other  surrounding comparison stars. We used the
first twenty  brightest stars in our  field of view,  and calculated a
robust  weighted average  of their  fluxes after  removing  any linear
differential  extinction  trend  as  measured  in  the  out-of-transit
segments of the  lightcurves.  Since the telescope has  a german mount
once it crosses the meridian it  flips around the field of view of 180
degrees. Once  this event happend we  found it was  necessary to apply
two distinct normalizations before  and after the meridian crossing in
order to match the photometric zero points.

\subsection{Time stamps}
\label{s:times}

We  report all  the  mid-exposure  times of  our  measurements to  the
Barycentric  Julian   Date  (BJD)  reference   frame  and  barycentric
dynamical time standard (TDB)  using the on-line converter provided by
Jason Eastman
\footnote{http://astroutils.astronomy.ohio-state.edu/time/utc2bjd.html}
(Eastman, Siverd \& Gaudi 2010).

\section{Stellar parameters}
\label{s:stellar_parameters}

We used a combined spectrum  of WASP-3 to derive spectroscopic stellar
atmospheric  parameters,  including   its  effective  temperature  and
metallicity.  The spectrum  used is a stacked of  8 individual spectra
obtained  between July and  August 2007  (Pollacco et  al.~2008).  The
spectra were downloaded from  the OHP-SOPHIE archive. All spectra were
obtained in  the HE mode (R$\sim$40\,000)  placing the fiber  B on the
sky.  We  used the spectrum in  fiber B to  subtract any contamination
light  in fiber  A  (pointing  to WASP-3),  after  correcting for  the
relative efficiency of  the two fibers.  The final  spectrum has a S/N
of the order of 100 in the 6500\,A region.

We  used  the  metholology   and  line-list  described  in  Santos  et
al.~(2004).  In  brief, after the  measurement of the  line equivalent
widths (EWs), the parameters are obtained making use of a line-list of
22 FeI  and 9  FeII lines and  forcing both excitation  and ionization
equilibrium.  We  refer to Santos  et al.  for details.   The analysis
was done in LTR using a  grid of Kurucz (1993) model atmospheres and a
recent version of the radiative transfer code MOOG Sneden (1973).  The
EWs were derived manually using the IRAF splot task.

The    final   obtained   stellar    parameters   are    as   follows:
T$_\mathrm{eff}$=6448$\pm$123\,K,         $\log{g}$=4.49$\pm$0.08\,dex,
$\xi_t$=2.01$\pm$0.40\,km\.s$^{-1}$,  and  [Fe/H]=-0.02$\pm$0.08\,dex.
As  a  double  check,  we  also independently  derived  the  effective
temperature of  the star using  the line-ratio procedure  described in
Sousa et  al.~(2010).  This procedure uses a  different line-list, and
the  EWs are  measured automatically  using  the {\sc  ares} (Sousa  et
al.~2007) code. The effective temperature derived using this method is
6432$\pm$94\,K,  in   perfect  agreement  with   the  value  mentioned
above. These  values are in agreement  with the ones  presented in the
planet  announcement  paper  (Pollacco  et al.~2008),  who  derived  a
temperature    of    6400$\pm$100\,K,    a    surface    gravity    of
4.25$\pm$0.05\,dex, and a metallicity of 0.00$\pm$0.20\,dex.

\section{Transit analysis}
\label{s:transit_analysis}

We modeled the observed transits considering the analytical formula of
Mandel  \&  Agol  (2002).   We  adopted in  particular  the  following
parametrization  for  the  planet   distance  to  the  stellar  center
normalized to the stellar radius ($z$):

\begin{eqnarray}
z^2(t)\, & = & \Big(\frac{8\,\pi^2\,\rm G}{3\,\rm P}\Big)^{2/3}\,\rho_{\star}^{2/3}\,\Big[(t-T_0)^2-\Big(\frac{T_d}{2}\Big)^2\Big]\,+ \nonumber \\
         & + & \,(1\,+\,r)^2
\end{eqnarray}

\noindent
where G is the gravitational constant,  P is the orbital period of the
planet, $\rho_{\star}$ is the mean  stellar density, $T_0$ is the time
of  transit minimum,  $T_d$ is  the total  transit duration  (from the
first to  the fourth contact)  and $r$ is  the ratio of  the planetary
radius  to  the stellar  radius.   We  fit  each lightcurve  with  the
Levemberg-Marquardt algorithm (Press 1992). We made use of the partial
derivatives of the flux loss  calculated by P\'al (2008) as a function
of the radius ratio $r$ and the normalized distance $z$.

The flux $F$ of the star at  each given instant of time $t$ during the
transit (corresponding to the  normalized distance $z$) was assumed to
be

\begin{equation}
F(z(t))\,=\,F_{MA}(z(t))\,+\,$A$\,\times\,\rm Air(t)\,+\,$B$,
\end{equation}

\noindent
where $F_{MA}$ is the flux loss predicted by the Mandel \& Agol (2002)
formula, Air(t) is the airmass at the instant $t$, and $A$ and $B$ are
two  parameters  to account  for  a  residual  photometric trend  with
airmass and  a constant zero  point offset.  We therefore  assumed six
free  parameters: the  time of  transit minimum  ($T_0$),  the airmass
coefficient ($A$),  the constant  zero point $B$,  the planet  to star
radius ratio ($r$), the transit  duration ($T_d$) and the mean stellar
density ($\rho_{\star}$).

We  assumed a  quadratic limb  darkening law  which  coefficients were
fixed  interpolating  the  tables  of  Claret  \&  Bloemen  (2011)  in
correpondence  to  the spectroscopic  parameters  of  the star.   This
procedure  yielded   the  following  coefficients   in  the  $I$-band:
$g_1=0.2150$ for  the linear term  and $g_2=0.3034$ for  the quadratic
term.  The orbital period of the planet was fixed as well at the value
of $P=1.846834$  days (Pollacco et al.~2008).  For  each iteration the
Levemberg-Marquardt algorithm  calculated the reduced  $\chi_{red}$ of
the fit defined as:

\begin{equation}
\chi_{red}\,=\,\sqrt{\sum_{i=1}^{i=N}\frac{(O_i-F_i)^2}{N-N_{free}}}
\end{equation}

\noindent
where  $O_i$   is  the  observed   flux  corresponding  to   the  i-th
measurement, $F_i$  is the model  calculated flux as  described above,
$N$ is the total number  of measurements, and $N_{free}$ the number of
free  parameters.  The  Levemberg-Marquardt algorithm  found  the best
solution  by  means of  $\chi_{red}$  minimization.  This solution  is
however  only  a  formal  solution,  the  best  parameters  and  their
uncertainties  were  then found  using  a  Markov  Chain algorithm  as
described below.

\subsection{Uncertainties of the observations}
\label{s:uncertainties}

\begin{figure}
\begin{center}
\includegraphics[width=7.5cm]{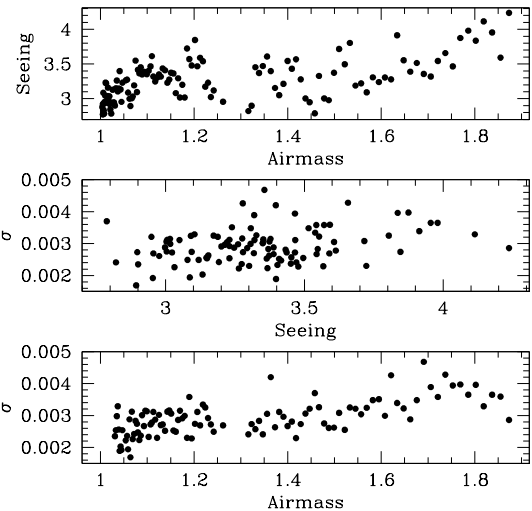}
\caption{
Noise correlation with atmospheric indicators relative to April 13, 2011.
}
\label{fig:noise}
\end{center}
\end{figure}

To  each measurement  in  our datasets  we  associated an  uncertainty
accordingly to the  photon noise and the read  out noise as determined
by {\sc  daophot}. These errors were  then added in  quadrature to the
scatter  of the  residual fluxes  of the  comparison stars  around our
derived mean  averaged values (see  Sect.~\ref{s:corrected}), and then
rescaled  in such  a  way that  the best  models  we fit  to the  data
produced  a $\chi^2$=1.   As pointed  out  by Pont  et al.~(2006)  the
presence of correlated noise in the data strongly limits the precision
of the  observations.  The  uncertainties calculated by  {\sc daophot}
already  accounted  for  some  obvious noise  correlations.   This  is
evident in  Fig.~\ref{fig:noise} where we plot the  uncertainty of the
measurements as a  function of airmass and seeing.   This ensured that
the  transits are  fitted  giving more  weight  to those  measurements
acquired  under the  best observing  conditions.  Nonetheless,  it may
well be that the noise in  our data is correlated also with some other
non trivial  variables that our  reduction did not take  into account.
In order to verify this hypothesis we created some mock lightcurves of
our  model-subtracted lightcurves assuming  that each  simulated point
was  distributed  normally  around  zero  but  with  a  time-dependent
dispersion equal to the uncertainty of the correspondent real data. We
then compared the  RMS of the real and  simulated lightcurves averaged
over timescales  comprised in between 10  min to 30  min.  The average
ratio  of  the   dispersions  of  the  real  to   the  simulated  data
($\sigma_r/\sigma_s$) was  always smaller than one  with the exception
of    epoch   574    ($\sigma_r/\sigma_s=1.09$),    and   epoch    627
($\sigma_r/\sigma_s=1.04$)  observations.   For  these two  nights  we
expanded our uncertainties by these factors, whereas for the remaining
nights we didn't apply any other correction.

\subsection{Markov Chain Monte Carlo analysis}
\label{s:markov}

A  commonly  used  approach   to  derive  parameter  uncertainties  in
exoplanetary literature  (e. g.   Gazak, J. et  al. 2012) is  based on
Markov Chain Monte Carlo analysis.   We implemented our own version of
the Markov Chain algorithm along the following lines. For each transit
lightcurve we  created five  chains of 10$^{5}$  steps. Each  chain is
started from  a point 5-$\sigma$  away (in one randomly  selected free
parameter)   from   the   best-fitting   solution  obtained   by   the
Levenberg-Marquardt  algorithm,  where  the  $\sigma$  values  of  the
parameters considered are those that  are obtained by the same transit
fitting algorithm as described in Sect.~\ref{s:transit_analysis}.  The
$\chi^2_{\rm old}$ of the fit of this initial solution is recorded and
compared  with  the  $\chi^2_{\rm  new}$  obtained  in  the  following
step. The following step is obtained jumping from the initial position
to  another  one  in  the multidimensional  parameter  space  randomly
selecting  one of the  free parameters  and changing  its value  by an
arbitrary  amount  which is  dependent  on  a  jump constant  and  the
uncertainty $\sigma$  of the parameter itself.  Steps  are accepted or
rejected  accordingly   to  the  Metropolis-Hastings   criterium.   If
$\chi^2_{\rm  new}$  is lower  than  $\chi^2_{\rm  old}$  the step  is
executed,      otherwise     the     execution      probability     is
$P=e^{-\Delta\chi^2/2}$         where        $\Delta\chi^2=\chi^2_{\rm
  new}-\chi^2_{\rm old}$.   In this  latter situation a  random number
between 0 and 1 is  drawn from a uniform probability distribution.  If
this number is lower than $P$ then the step is executed, otherwise the
step  is rejected and  the previous  step is  repeated instead  in the
chain.  In  any case  the value of  the $\chi^2$  of the last  step is
recorded and compared with the one of the following step up to the end
of the chain.  We adjusted the jump constants (one for each parameter)
in such a way that the step acceptance rate for all the parameters was
around 25  per cent.  The  convergence among the five  separate chains
was checked  comparing the variances within and  between the different
chains by means of the Gelman \& Rubin (1992) statistic.  In all cases
the values of the Gelman-Rubin statistic was within a few percent from
unity indicating  that the chains  were converged and well  mixed.  We
then excluded the  first 20 per cent steps of each  chain to avoid the
initial burn-in phase, and for  each parameter we merged the remaining
part  of  the  chains together.   Then  we  derived  the mode  of  the
resulting  distributions,  and the  68.3  per  cent confidence  limits
defined by the  15.85th and the 84.15th percentiles  in the cumulative
distributions.

We  run  two separate  groups  of  chains  first considering  as  free
parameters $T_0$, $A$ and $B$  while retaining the others fixed at the
best values obtained by the Levemberg-Marquardt algorithm. In a second
step we instead perturbed $\rho_{\star}$, $T_d$ and $r$ while retaining
the remaining  parameters fixed  at the values  obtained in  the first
step.   Perturbing all  the  parameters together  lead  in general  to
unstable  convergence in  particular for  the airmass  and  zero point
coefficients so we decided to split the procedure in two steps.

\subsection{Mean stellar density}
\label{s:stellar_density}

The mean stellar density is  one of the most important parameters that
can  be  extracted from  transiting  planet  lightcurves (Sozzetti  et
al.~2007). It  is interesting to  compare the stellar  density derived
from the  analysis of our lightcurves  to the value  obtained from the
analysis of other datasets.

We then considered the  precise transit lightcurve of WASP-3b obtained
by Tripathi  et al.~(2010) in the Sloan  z$^{\prime}$-band filter with
the  University  of Hawaii  2.2m  Telescope.   We  chose this  transit
because all  the other transits of  WASP-3b published so  far (both by
Tripathi  and  by  other  authors)  have been  observed  with  smaller
telescopes, and because  being the observations carried out  in a near
infrared  filter the  impact of  limb-darkening on  the  transit shape
should be  lower than  at shorter wavelenghts.   We performed  on this
lightcurve the  fit and the statistical  analysis previously described
obtaining  final values for  the parameters  consistent with  those of
Tripathi  et al.~(2010).   For the  mean stellar  density  we obtained
$\rho_{\star}\,=\,0.50^{+0.15}_{-0.06}$  g$\,  \rm  cm^{-3}$.  On  the
contrary  the analysis  of our  lightcurves favours  a  larger density
equal  to $\rho_{\star}\,=\,(0.80\,\pm\,0.07)$  g$\,  \rm cm^{-3}$  by
taking   the  mean   average  of   the  results   reported   in  Table
\ref{tab:results}. We  note that while these  estimates are consistent
within  2$\sigma$ the  value  obtained from  our  lightcurves is  more
similar   to   the   one   reported   by   Pollacco   et   al.~(2008),
$\rho_{\star}\,=\,(0.55^{+0.15}_{−0.05})\,\rho_{\odot}\,=\,0.77^{+0.21}_{-0.07}$
g$\,      \rm     cm^{-3}$      and      Miller     et      al.~(2010)
$\rho_{\star}\,=\,(0.67^{+0.05}_{−0.06})\,\rho_{\odot}\,=\,0.94^{+0.07}_{-0.08}$
g$\, \rm cm^{-3}$.

\subsection{Results}
\label{s:results}

In Table~\ref{tab:results}  we reported for each  transit our measured
transit  durations ($\rm  T_d$) and  planet to  stellar  radius ratios
($\rm  r$).  In  particular  we observed  a  weighted average  transit
duration  $\rm T_d$  equal to  $\rm T_d=(158\,\pm\,1)$  min,  which is
closer  to   the  value  reported  by  Pollacco   et  al.~(2008)  $\rm
T_d=(159.8^{+1.3}_{-2.6})$  min  and  Maciejewski et  al.~(2010)  $\rm
T_d=(161.2\,\pm\,2.3)$  min,  than   those  reported  by  Tripathi  et
al.~(2010), $\rm  T_d=(168.8\,\pm\,0.7)$ min and  Gibson et al.~(2008)
$\rm T_d=(165.2^{+1.2}_{-0.8})$ min.

Our weighted average  planet to stellar radius ratio  is equal to $\rm
r=(0.1061 \pm 0.0007)$ and it is consistent with the value reported by
Maciejewski et al.~(2010)  $\rm r =(0.108 \pm 0.003)$  and Tripathi et
al.      (2010)      for     the     $z^{\prime}$      filter     $\rm
r=0.1099^{+0.0006}_{-0.0010}$, but it is  larger than the values given
by  Pollacco et  al.~(2008) $\rm  r =  0.1030^{+0.0010}_{-0.0015}$ and
Gibson et al.~(2008) $\rm r = 0.1014^{+0.0010}_{-0.0008}$.

Table~\ref{tab:midtransits} lists the collection of transit timings of
WASP-3b   presented  in   published   papers,  along   with  our   new
measurements.  Our timing errors are  comprised between 80 sec and 233
sec.   Note  that  Maciejewski   et  al.~(2010)  transit  timings  are
expressed in BJD based on  TT (Terrestrial Time).  The difference with
respect to  BJD based  on TDB is  however negligible for  our purposes
(Eastman,  Siverd \&  Gaudi  2010).  Transit  timings  of Pollacco  et
al.~(2008), Tripathi et al.~(2010)  and Gibson et al.~(2008) have been
corrected  to  account  for   the  conversion  between  HJD  and  $\rm
BJD_{TDB}$\footnote{   Jason   Eastman's   Barycentric   Julian   Date
  Converter, http://astroutils.astronomy.ohio-state.edu/time/}.

\noindent
We   considered    all   the   transit    ephemerides   presented   in
Table~\ref{tab:midtransits}  and recalculated  the  transit period  by
fitting  a  weighted  linear  least  square  model  to  all  the  data
obtaining:

\begin{eqnarray}
T_C(E) & = & \,(2454605.5601\,\pm\,0.0002)\,+ \nonumber \\
       & + & E\,\times\,(1.846834\,\pm\,0.000001)
\end{eqnarray}

\noindent
where $E$ is the transit epoch.  We then subtracted the model from the
observed  ephemerides  which gave  the  (O-C)  residuals presented  in
Table~\ref{tab:midtransits}    and    in   Fig.~\ref{fig:ttv}.     Our
measurements are  consistent with the calculated  ephemerides with the
exception  of those  relative  to  epochs 196  and  574.  The  reduced
chi-squared  value of the  fit ($\sqrt{\chi^2_r}$)  is equal  to 2.30,
obtained    from    all    the    40    measurements    reported    in
Table~\ref{tab:midtransits} and considering 2 degrees of freedom.

\begin{figure*}
\begin{center}
\includegraphics[width=7.3cm]{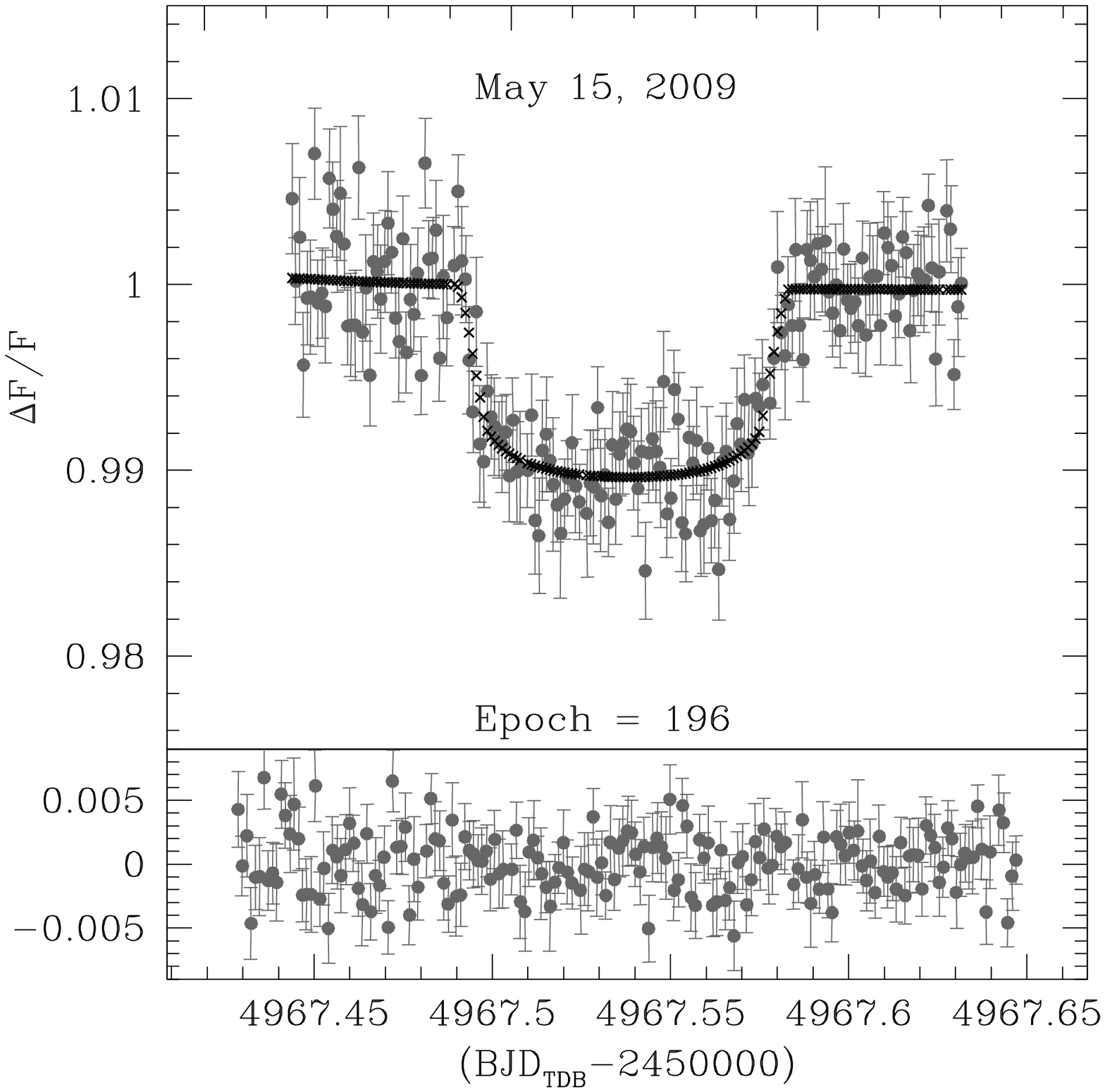}
\includegraphics[width=7.3cm]{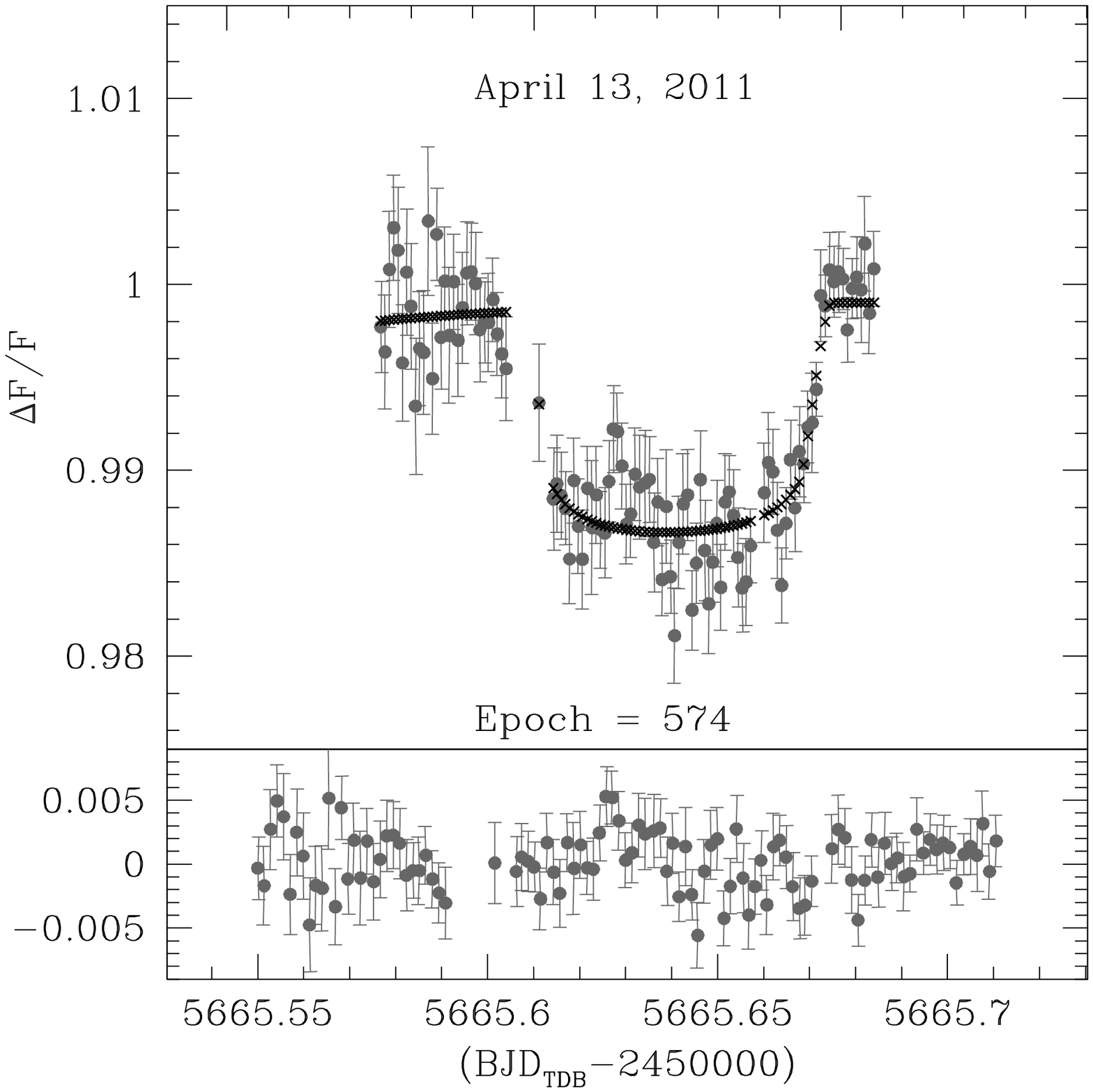}
\includegraphics[width=7.3cm]{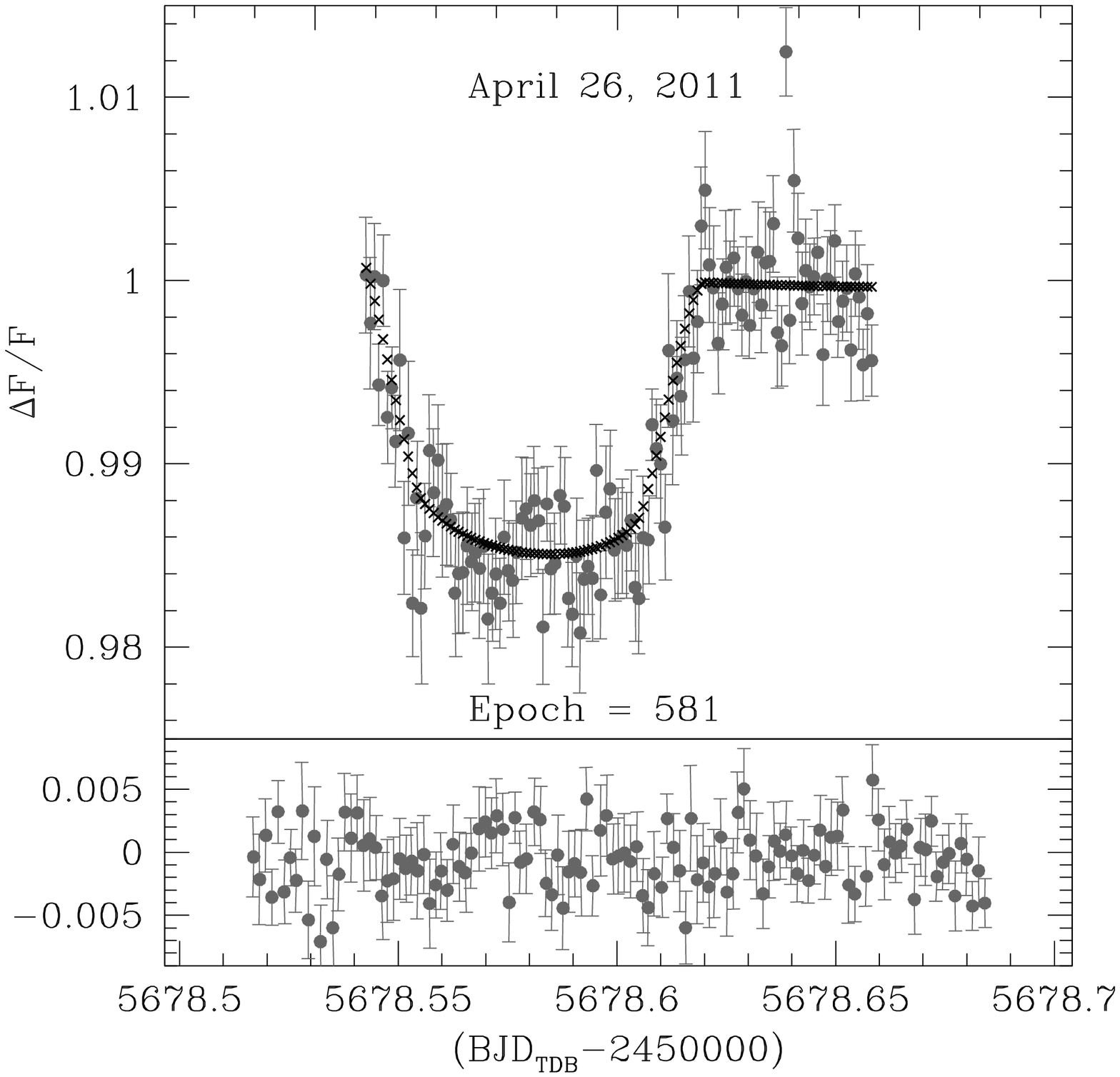}
\includegraphics[width=7.3cm]{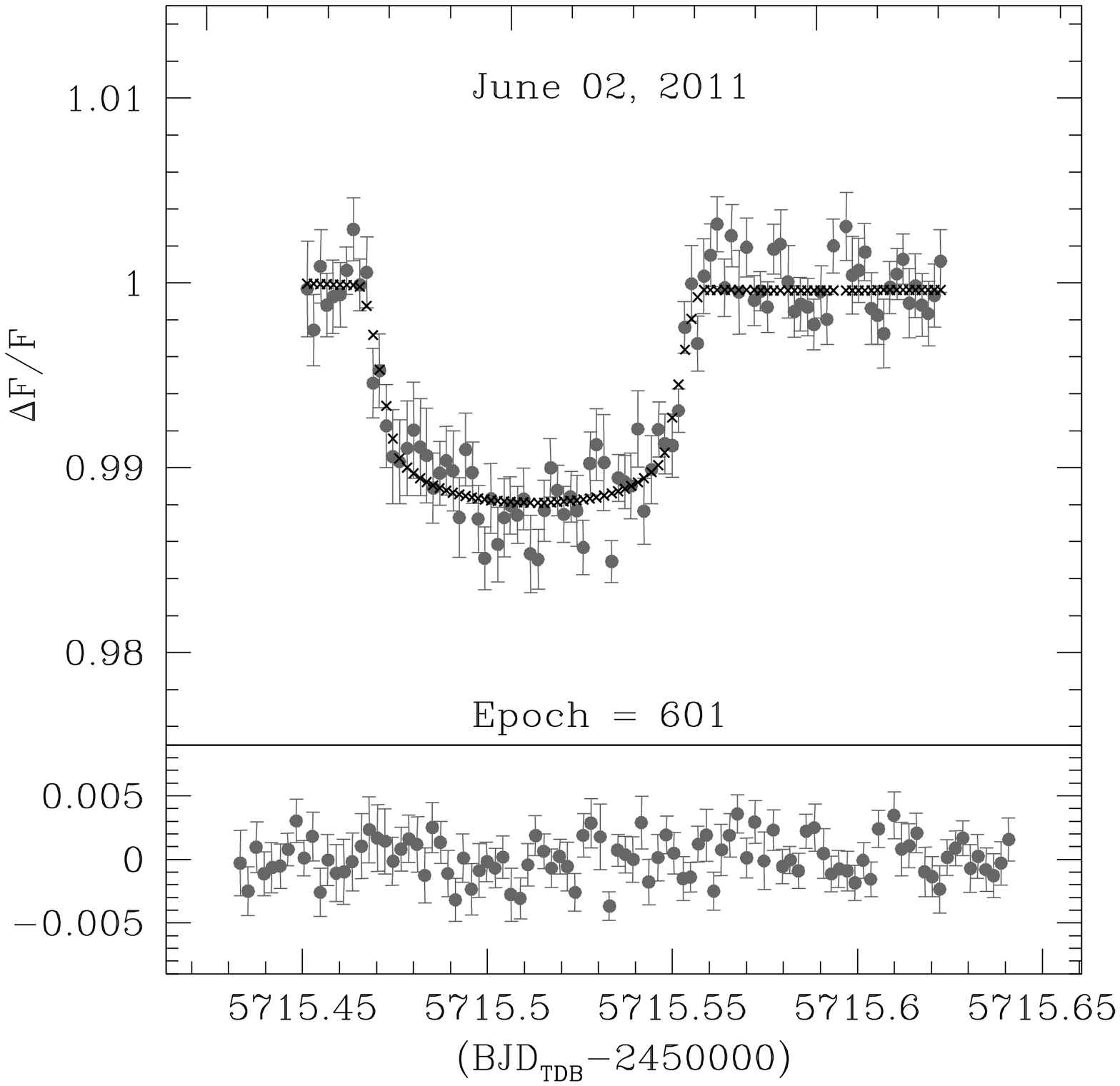}
\includegraphics[width=7.3cm]{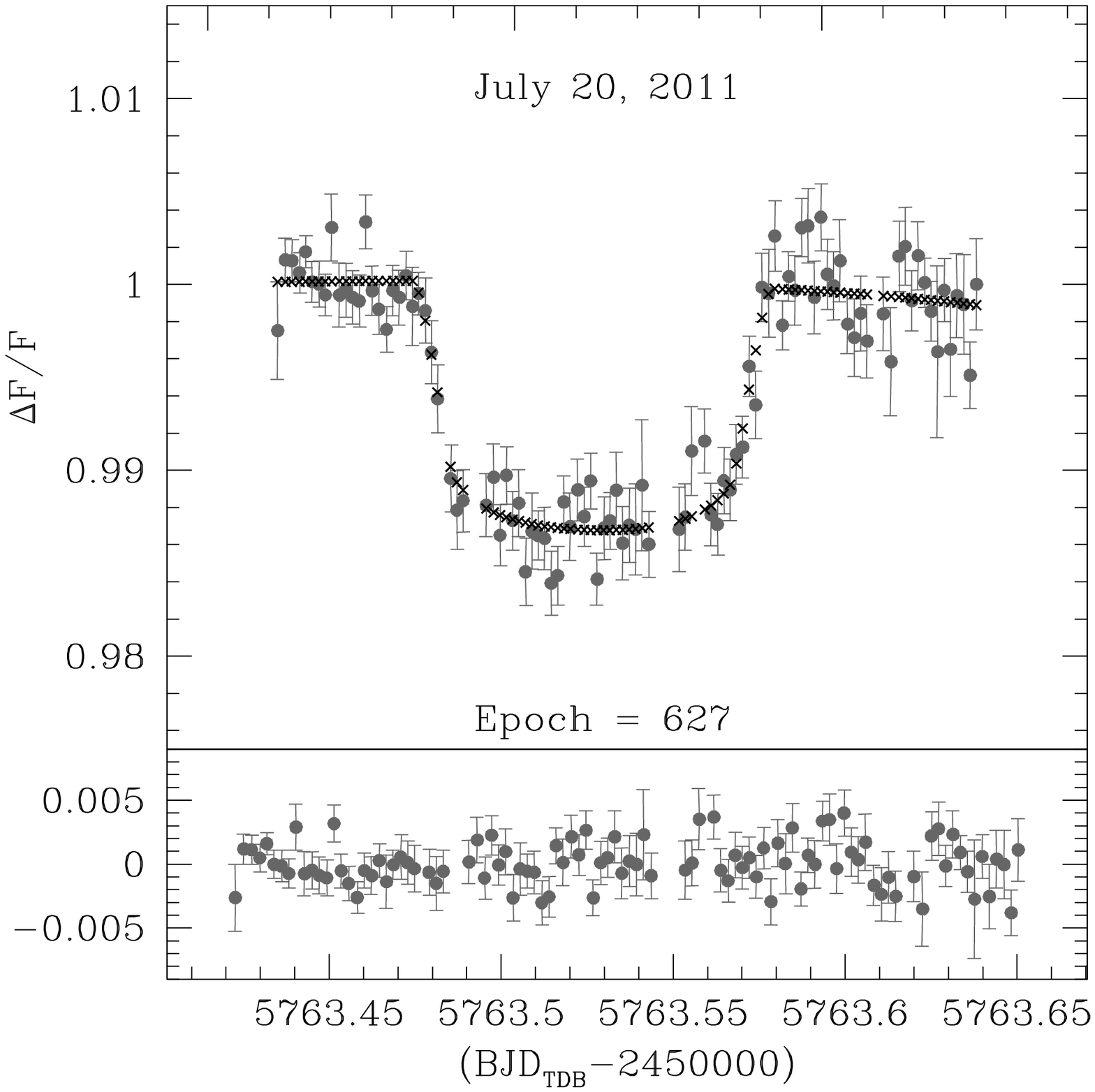}
\includegraphics[width=7.3cm]{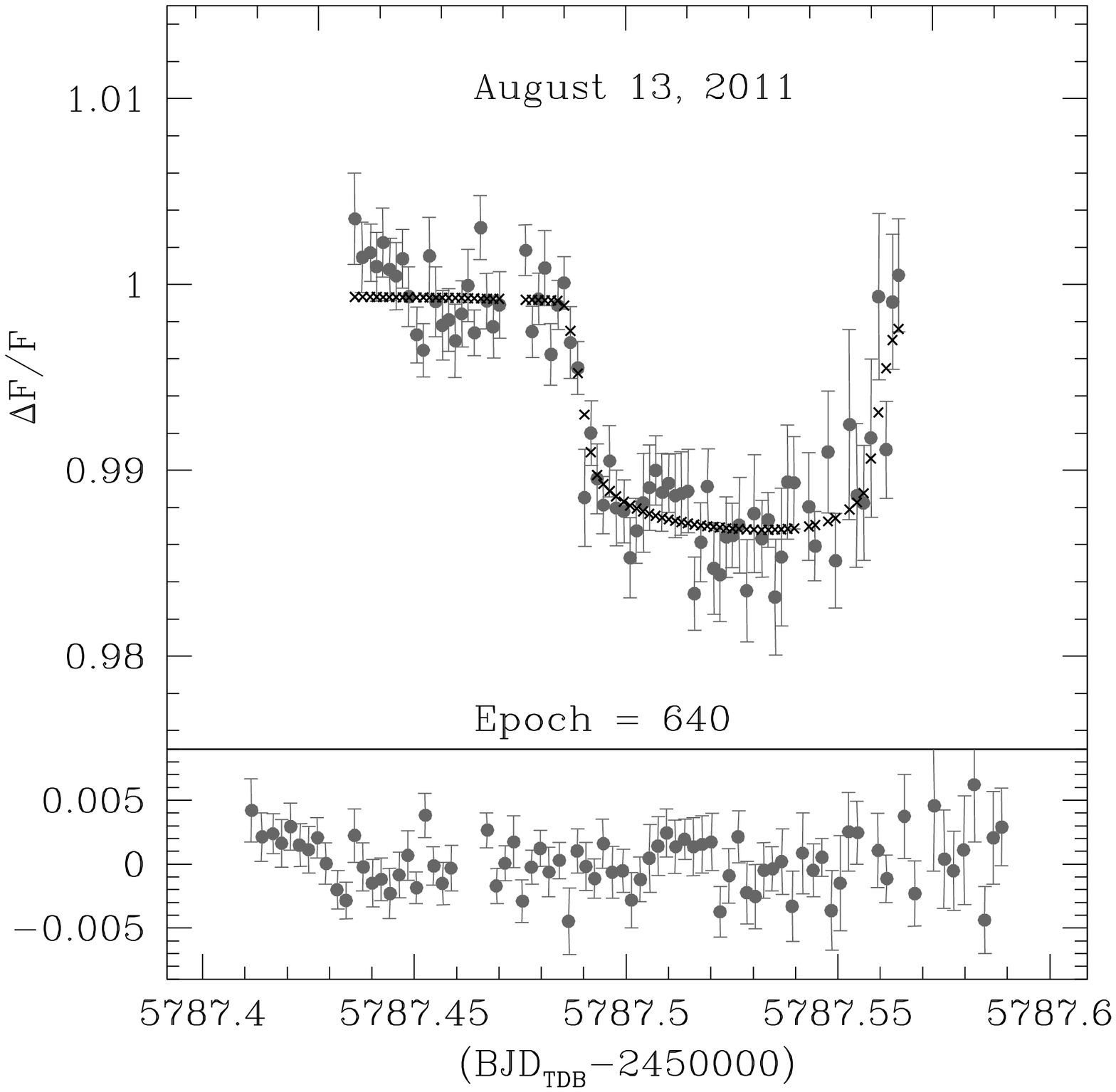}
\caption{
Observed transits of WASP-3b along with our best-fitting models and residuals.
}
\label{fig:transits}
\end{center}
\end{figure*}

\addtocounter{figure}{-1}

\begin{figure*}
\begin{center}
\includegraphics[width=7.3cm]{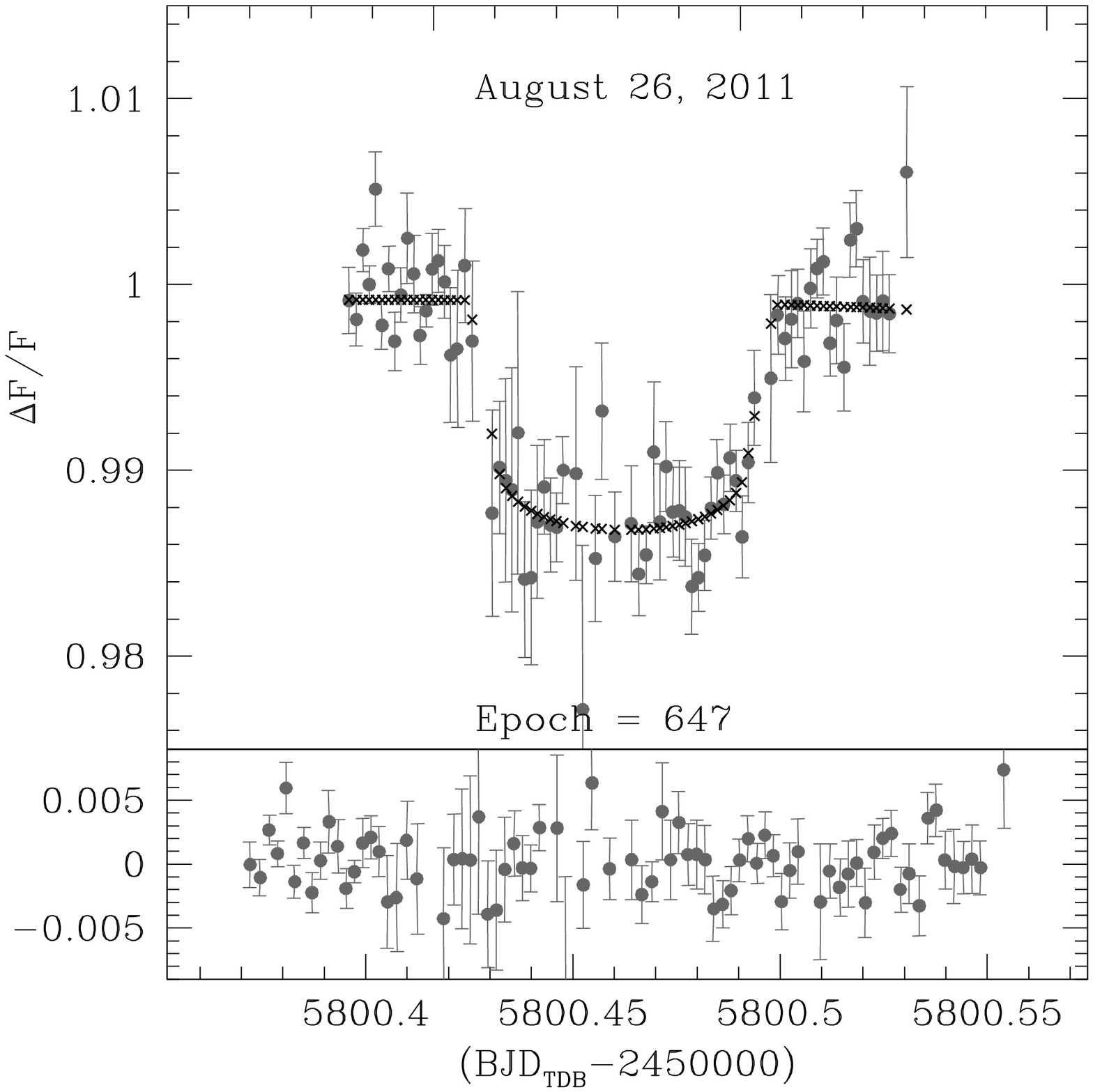}
\includegraphics[width=7.3cm]{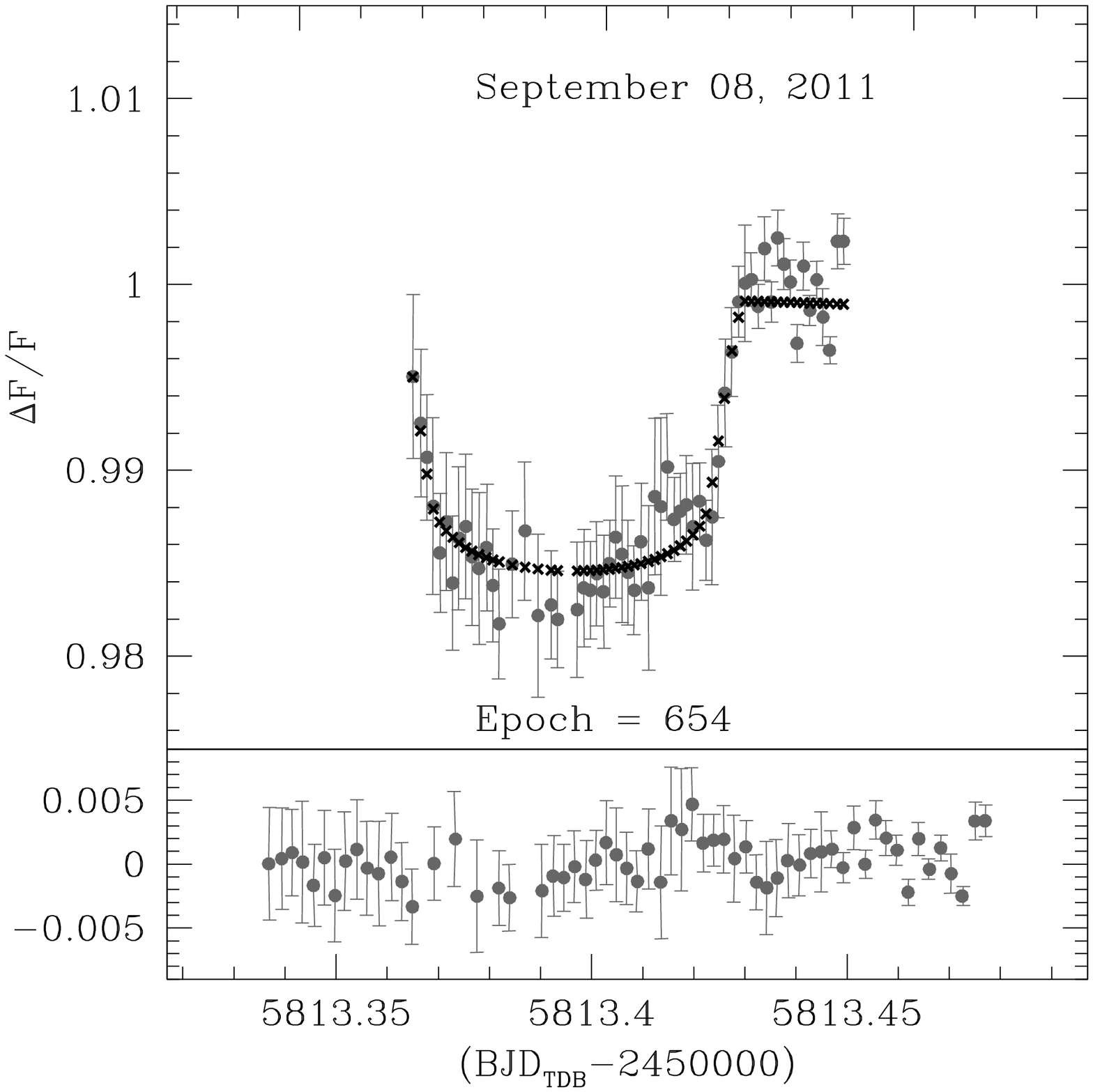}
\caption{
-- {\it Continued}
}
\label{fig:transits}
\end{center}
\end{figure*}

\noindent
In order to further check  our ephemerides measurements we applied the
barycentric method (Szab\'{o} et al.  2006, Oshagh et al.~2012).  This
technique  calculates  the transit  center  ($\rm  T_C$)  as the  flux
weighted average epoch across the transit:

\begin{equation}
T_C\,=\,\frac{\sum_{i=1}^{i=N}\,t_i\,(1\,-\,f_i)}{\sum_{i=1}^{i=N}\,(1\,-\,f_i)}
\end{equation}

\noindent
where $f_i$  is the normalized  flux, $t_i$ the  time and N  the total
number  of  measurements  within  one transit.   Transit  fitting  and
barycentric  method  results  agree  within  1.9-$\sigma$  (where  the
$\sigma$   is   the  average   of   the   uncertainties  reported   in
Table~\ref{tab:midtransits}) for  all the transits  with the exception
of the transit occurring at epoch 640 (3.7-$\sigma$).  In any case, we
notice that this  transit is almost a partial transit  which is at the
limit of applicability of the barycentric method.

\begin{table*}
\caption{
Best-fitting transit parameters.
\label{tab:results}
}
\begin{center}
\resizebox{17.0cm}{!}{
\begin{tabular}{c c c c c c c}
\hline
Date & Epoch ($E$) & Slope ($s$) & Constant ($c$) & $\rm \frac{R_p}{R_s}$ ($\rm r$) & Duration ($\rm T_d$) &        Mean stellar density ($\rho_{\star}$)       \\
     &             &             &                &                                 &       (min)          &         (g$\rm \,cm^{-3}$)      \\
\hline
15/05/2009 & 196 & $ 0.00060_{-0.00060}^{+0.00074}$ & $0.99913_{-0.00094}^{+0.00077}$ &  $0.09506_{-0.00123}^{+0.00309}$ &  $156.4_{-2.3}^{+4.7} $ & $1.05740_{-0.42994}^{+0.02047}$ \\
           &     &                               &                              &                               &                        &                              \\
13/04/2011 & 574 & $-0.00164_{-0.00172}^{+0.00115}$ & $1.00068_{-0.00135}^{+0.00203}$ &  $0.10383_{-0.00110}^{+0.00256}$ &  $145.4_{-2.6}^{+3.2} $ & $1.37746_{-0.45132}^{+0.02257}$ \\
           &     &                               &                              &                               &                        &                              \\
26/04/2011 & 581 & $ 0.00393_{-0.00268}^{+0.00115}$ & $0.99571_{-0.00225}^{+0.00225}$ &  $0.12177_{-0.00319}^{+0.00172}$ &  $166.0_{-5.0}^{+5.0} $ & $0.51945_{-0.06303}^{+0.25213}$ \\
           &     &                               &                              &                               &                        &                              \\
02/06/2011 & 601 & $ 0.00078_{-0.00185}^{+0.00123}$ & $0.99882_{-0.00153}^{+0.00187}$ &  $0.10226_{-0.00134}^{+0.00269}$ &  $162.1_{-2.5}^{+5.0} $ & $0.80799_{-0.24422}^{+0.16908}$ \\
           &     &                               &                              &                               &                        &                              \\
20/07/2011 & 627 & $-0.00223_{-0.00106}^{+0.00106}$ & $1.00243_{-0.00142}^{+0.00095}$ &  $0.10904_{-0.00192}^{+0.00128}$ &  $168.1_{-2.3}^{+3.8} $ & $0.82458_{-0.20157}^{+0.09406}$ \\
           &     &                               &                              &                               &                        &                              \\
13/08/2011 & 640 & $-0.00274_{-0.00163}^{+0.00163}$ & $1.00208_{-0.00223}^{+0.00149}$ &  $0.10222_{-0.00166}^{+0.00249}$ &  $157.8_{-2.0}^{+5.1} $ & $1.01379_{-0.35486}^{+0.03735}$ \\
           &     &                               &                              &                               &                        &                              \\
26/08/2011 & 647 & $-0.00080_{-0.00137}^{+0.00150}$ & $0.99998_{-0.00175}^{+0.00143}$ &  $0.10534_{-0.00293}^{+0.00267}$ &  $147.0_{-4.2}^{+10.4}$ & $1.09265_{-0.48430}^{+0.22601}$ \\
           &     &                               &                              &                               &                        &                              \\
08/09/2011 & 654 & $-0.00122_{-0.00308}^{+0.00205}$ & $1.00056_{-0.00312}^{+0.00312}$ &  $0.11422_{-0.00203}^{+0.00304}$ &  $163.8_{-3.3}^{+5.4} $ & $0.95100_{-0.31570}^{+0.05262}$ \\
           &     &                               &                              &                               &                        &                              \\
\hline
\end{tabular}
}
\end{center}
\end{table*}

\begin{table*}
\caption{
Collection of transit timing measurements of WASP-3b in chronological order.
\label{tab:midtransits}
}
\begin{center}
\resizebox{17.0cm}{!}{
\begin{tabular}{c c c c c c c c}
\hline
Epoch & Time of transit minimum &  $\Delta^{-}(\rm BJD_{TDB})$    &  $\Delta^{+}(\rm BJD_{TDB})$ &   $\rm (O\,-\,C)$  &  $\Delta^{-}(\rm O\,-\,C)$ &  $\Delta^{+}(\rm O\,-\,C)$  & Reference\\
      & ($\rm BJD_{TDB}$-2450000)  &      (days)         &   (days)         &       (sec)    &        (sec)          &         (sec)           &          \\
\hline
      -250.   &      4143.85104     &       0.00040      &      0.00040    &    -46.    &     35.   &      35.  &  Pollacco et al.~(2008) \\              
        -2.   &      4601.86588     &       0.00027      &      0.00027    &    -44.    &     23.   &      23.  &  Tripathi et al.~(2010) \\              
         0.   &      4605.56030     &       0.00035      &      0.00035    &     21.    &     30.   &      30.  &  Gibson et al.~(2008) \\                
        12.   &      4627.72172     &       0.00031      &      0.00031    &    -30.    &     27.   &      27.  &  Tripathi et al.~(2010) \\              
        18.   &      4638.80403     &       0.00031      &      0.00031    &     83.    &     27.   &      27.  &  Tripathi et al.~(2010) \\              
        30.   &      4660.96509     &       0.00021      &      0.00021    &      1.    &     18.   &      18.  &  Tripathi et al.~(2010) \\              
        40.   &      4679.43269     &       0.00050      &      0.00050    &    -63.    &     43.   &      43.  &  Christiansen et. al.~(2011) \\         
        41.   &      4681.27911     &       0.00040      &      0.00040    &    -99.    &     35.   &      35.  &  Christiansen et. al.~(2011) \\         
        42.   &      4683.12740     &       0.00035      &      0.00035    &     27.    &     30.   &      30.  &  Christiansen et. al.~(2011) \\         
        43.   &      4684.97486     &       0.00027      &      0.00027    &     81.    &     23.   &      23.  &  Christiansen et. al.~(2011) \\         
        44.   &      4686.82053     &       0.00059      &      0.00059    &    -19.    &     51.   &      51.  &  Christiansen et. al.~(2011) \\         
        46.   &      4690.51381     &       0.00055      &      0.00055    &    -53.    &     48.   &      48.  &  Christiansen et. al.~(2011) \\         
        47.   &      4692.36117     &       0.00043      &      0.00043    &     -7.    &     37.   &      37.  &  Christiansen et. al.~(2011) \\         
        48.   &      4694.20711     &       0.00042      &      0.00042    &    -84.    &     36.   &      36.  &  Christiansen et. al.~(2011) \\         
        59.   &      4714.52284     &       0.00036      &      0.00036    &    -36.    &     31.   &      31.  &  Gibson et al.~(2008) \\                
       194.   &      4963.84436     &       0.00072      &      0.00072    &   -128.    &     62.   &      62.  &  Tripathi et al.~(2010) \\              
       194.   &      4963.84563     &       0.00055      &      0.00055    &    -18.    &     48.   &      48.  &  Sada et al.~(2012) \\                  
       196.   &      4967.53651     &       0.00057      &      0.00085    &   -259.    &     49.   &      73.  &     This work \\                        
       201.   &      4976.77365     &       0.00051      &      0.00051    &     -3.    &     44.   &      44.  &  Tripathi et al.~(2010) \\              
       236.   &      5041.41271     &       0.00049      &      0.00049    &    -14.    &     42.   &      42.  &  Maciejewski et al.~(2010) \\           
       249.   &      5065.41995     &       0.00059      &      0.00059    &   -152.    &     51.   &      51.  &  Maciejewski et al.~(2010) \\           
       256.   &      5078.34873     &       0.00058      &      0.00058    &    -71.    &     50.   &      50.  &  Maciejewski et al.~(2010) \\           
       269.   &      5102.35933     &       0.00056      &      0.00056    &     81.    &     48.   &      48.  &  Maciejewski et al.~(2010) \\           
       289.   &      5139.29713     &       0.00049      &      0.00049    &    178.    &     42.   &      42.  &  Maciejewski et al.~(2010) \\           
       379.   &      5305.51082     &       0.00039      &      0.00039    &     60.    &     34.   &      34.  &  Maciejewski et al.~(2010) \\           
       403.   &      5349.83457     &       0.00039      &      0.00039    &     37.    &     34.   &      34.  &  Sada et al.~(2012) \\                  
       403.   &      5349.83182     &       0.00039      &      0.00039    &   -200.    &     34.   &      34.  &  Sada et al.~(2012) \\                  
       404.   &      5351.68320     &       0.00110      &      0.00110    &    192.    &     95.   &      95.  &  Littlefield~(2011) \\                  
       430.   &      5399.69990     &       0.00150      &      0.00150    &    108.    &    130.   &     130.  &  Littlefield~(2011) \\                  
       436.   &      5410.78020     &       0.00130      &      0.00130    &     47.    &    112.   &     112.  &  Littlefield~(2011) \\                  
       450.   &      5436.63590     &       0.00080      &      0.00080    &     49.    &     69.   &      69.  &  Littlefield~(2011) \\                  
       456.   &      5447.71550     &       0.00080      &      0.00080    &    -72.    &     69.   &      69.  &  Littlefield~(2011) \\                  
       574.   &      5665.64627     &       0.00069      &      0.00056    &    305.    &     60.   &      48.  &      This work \\                       
       581.   &      5678.57065     &       0.00106      &      0.00087    &      6.    &     92.   &      75.  &      This work \\                       
       592.   &      5698.88358     &       0.00060      &      0.00060    &   -188.    &     52.   &      52.  &  Sada et al.~(2012) \\                  
       601.   &      5715.50608     &       0.00074      &      0.00060    &   -102.    &     64.   &      52.  &      This work \\                       
       627.   &      5763.52552     &       0.00070      &      0.00047    &     50.    &     60.   &      41.  &      This work \\                       
       640.   &      5787.53379     &       0.00080      &      0.00080    &      1.    &     69.   &      69.  &      This work \\                       
       647.   &      5800.46112     &       0.00113      &      0.00170    &    -43.    &     98.   &     147.  &      This work \\                       
       654.   &      5813.38792     &       0.00098      &      0.00080    &   -133.    &     85.   &      69.  &      This work \\                       
\hline
\end{tabular}
}
\end{center}
\end{table*}

\section{Radial velocities}
\label{s:RV}

Radial velocities can be  used together with transit timing variations
to place  more stringent  constrains on the  presence of  a perturbing
object in  the WASP-3b system.   We therefore gathered all  the radial
velocity  measurements   of  WASP-3b publically   available  from  the
Exoplanet  Orbit  database\footnote{  http://exoplanets.org/}.   These
measurements  were presented  in Pollacco  et al.~(2008),  Simpson et
al.~(2010) and  Tripathi et al.~(2010)  and in the following  we first
introduced in more detail these datasets.

\subsection{Available data sets}

Pollacco et  al~(2008) obtained seven radial  velocity measurements of
WASP-3  using  the SOPHIE  spectrograph  at  the  1.93-m telescope  at
Haute-Provence  Observatory. The  observations were  performed between
2007 July  2-5 and August  27-30.  All the measurements  were acquired
outside the transit.
Simpson et al.~(2010) acquired 26 spectra of WASP-3 during the transit
occurring on  the night  of 2008 September  30. The  observations were
also obtained with the SOPHIE  spectrograph at the 1.93-m telescope at
Haute-Provence Observatory.   These authors also  reanalysed the seven
measurements presented  in Pollacco et al.~(2008) based  on an updated
version of the SOPHIE pipeline. We therefore decided to use these data
in  our study  and  not the  original  data presented  in Pollacco  et
al.~(2008).

\begin{figure}
\begin{center}
\includegraphics[width=8.0cm]{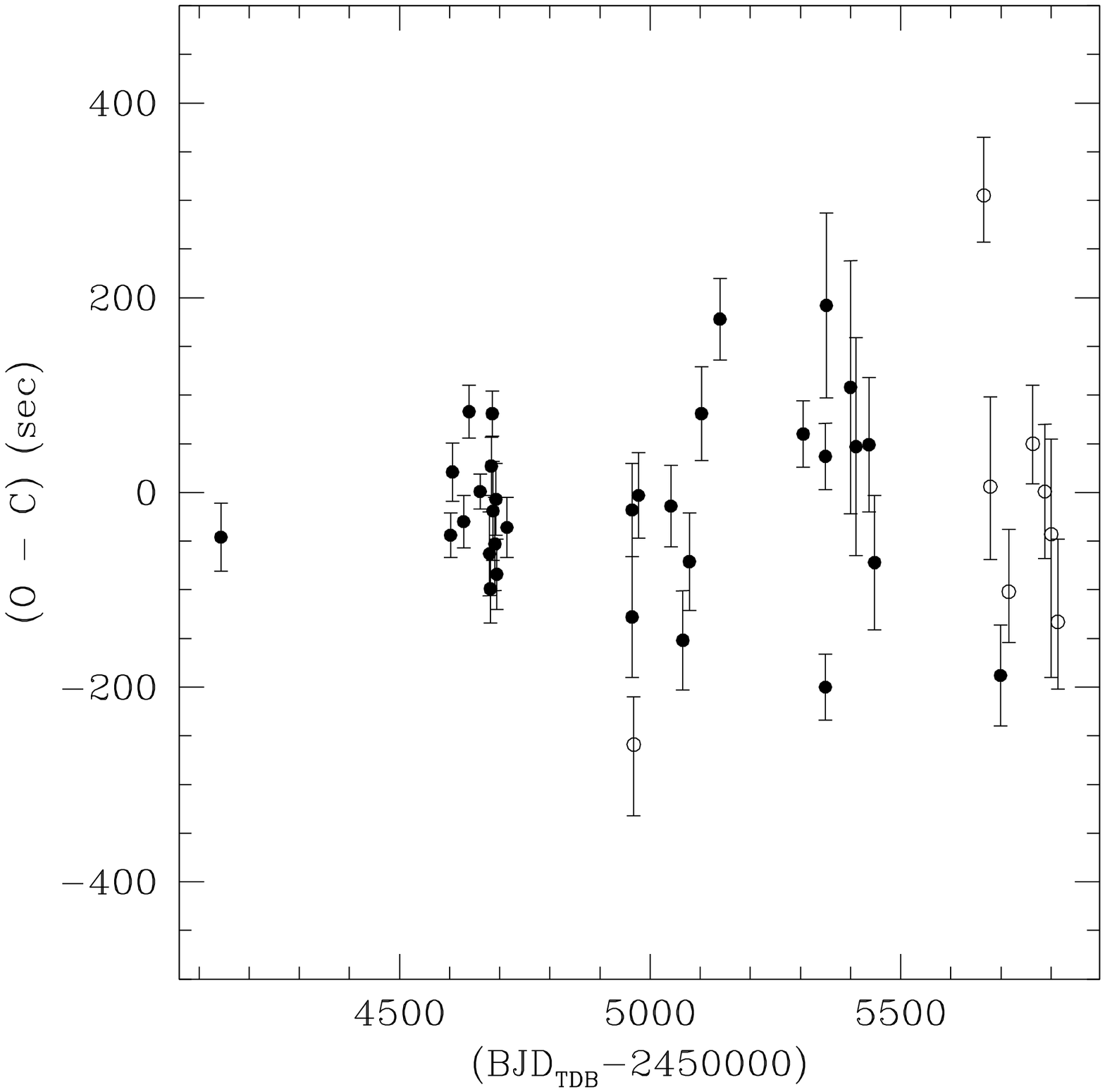}
\caption{
Observed  minus calculated  times of  transit minimum.  Filled circles
denote previous literature results,  open circles the new measurements
presented in this work.
}
\label{fig:ttv}
\end{center}
\end{figure}

Tripathi  et al.~(2010)  obtained 33  radial velocity  measurements of
WASP-3b  with  the  High   Resolution  Spectrometer  (HIRES;  Vogt  et
al.~1994) on the  Keck I 10 m telescope at the  W. M. Keck Observatory
on Mauna Kea.  The observations  were acquired both during the transit
(on 2008 June  19, 21 and on  2009 June 3) and outside  the transit on
several other nights  in 2008 and 2009. 

\section{Analysis of the radial velocity data}
\label{s:analysis_RV}

We  analyzed  simultaneously  all  the  radial  velocity  measurements
presented   both   by   Tripathi   et  al.~(2010)   and   Simpson   et
al.~(2010). Since  many measurements were acquired during the transit,
we fit the  data with a model describing both  the Keplerian motion of
the host  star and  the Rossiter-McLaughlin anomaly.  On one  hand we
modeled the Keplerian motion as:

\begin{equation}
RV\, = \, \tilde{K} \frac{\cos u +k}{\sqrt{1-h^2-k^2}} + \gamma 
\end{equation}

%
%
%
\noindent
where $\tilde{K}$  is the  radial velocity semi-amplitude  without the
contribution of the eccentricity  $e$, $k=e\,\cos \omega $, $h=e\,\sin
\omega   $,  $\gamma$   is   the  barycentric   radial  velocity   and
$u=\nu+\omega$ is  the true argument  of latitude with $\nu$  the true
anomaly and $\omega$ the argument of the pericenter.

On  the other  hand  we accounted  for  the Rossiter-McLaughlin  (RM)
anomaly  following  Hirano et  al.~(2010)  but  including an  improved
treatment of the RM effect during the partial phases of the transit (as
detailed  in     Appendix B):

\begin{equation}
RV_{RM}\, = -df\,\times\,v_p\,\Big[p-q\,\Big(\frac{v_p}{v\sin i }\Big)^2\Big] 
\end{equation}

\noindent
where $df$ is the flux loss due  to the transit of the planet in front
of the disk of the star, which we modeled as in Mandel \& Agol (2002),
$p$  and  $q$ are  two  parameters  related  to modellization  of  the
Rossiter-McLaughlin effect  as proposed  by Hirano et  al.~(2010) and
were fixed to the values adopted by Tripathi et al.~(2010) and Simpson
et  al.~(2010) as  reported  in Table~\ref{tab:RM_par},  $v_p$ is  the
average velocity  of the  star below the  area occulted by  the planet
(Hirano  et  al.  2010  and  Appendix B),  and  $v\sin i $  is  the
rotation velocity of the star.

\begin{table}
\caption{
Adopted values for the $p$ and $q$ parameters and for the limb darkening
coefficients entering in the Rossiter-McLaughlin model.
\label{tab:RM_par}
}
\begin{center}
\begin{tabular}{c c c c c}
\hline
\hline
$p$ & $q$  & $g_1$ & $g_2$ & Reference\\
\hline
1.51 & 0.44    & 0.596 & 0.215 & Tripathi et al.~(2010) \\
1.72 & 0.00546 & 0.69  & 0.    & Simpson et al.~(2010) \\
\hline
\end{tabular}
\end{center}
\end{table}

We  considered as  free parameters:  $\tilde{K}$, $h$,  $k$, $\lambda$
(the spin-orbit  angle), $ v\sin  i$, and $\gamma$.  Both  Tripathi et
al.~(2010)  and  Simpson  et  al.~(2010) distinguished  two  different
groups of data in their own dataset to account for possible systematic
radial velocity variations during their observing runs.  We decided to
perform the fit  twice, first following the analysis  of those authors
and  therefore   allowing  for  a  total  number   of  four  different
barycentric  radial velocities.   Then, we  redid the  fit considering
only two  different barycentric radial velocities for  the Tripathi et
al.~(2010) and  Simpson et al.~(2010) datasets.   This second approach
was intended to  check for possible long-term variations  among the RV
residuals  that could  have been  canceled out  by the  adoption  of a
larger number  of free  parameters.  Then, in  the end,  we considered
either nine  or seven free  parameters for the two  fits respectively.
We notice that Tripathi et al.~(2010) excluded from the fit three data
points which  were presenting a  clearly deviant radial  velocity with
respect to  the remaining measurements.  This  radial velocity $spike$
was  ultimately  attributed  by  the  authors  to  residual  moonlight
unexpectedly leaking into the  spectrograph and therefore we neglected
them  hereafter.    Additionally  Tripathi  et   al.~(2010)  added  in
quadrature  to  the uncertainties  of  their  data  a value  equal  to
14.8$\rm \,m\,s^{-1}$ to account for jitter noise.  Since, however, it
is not clear  which is the origin of this noise,  we didn't apply this
correction.

\begin{figure*}
\begin{center}
\includegraphics[width=7.2cm]{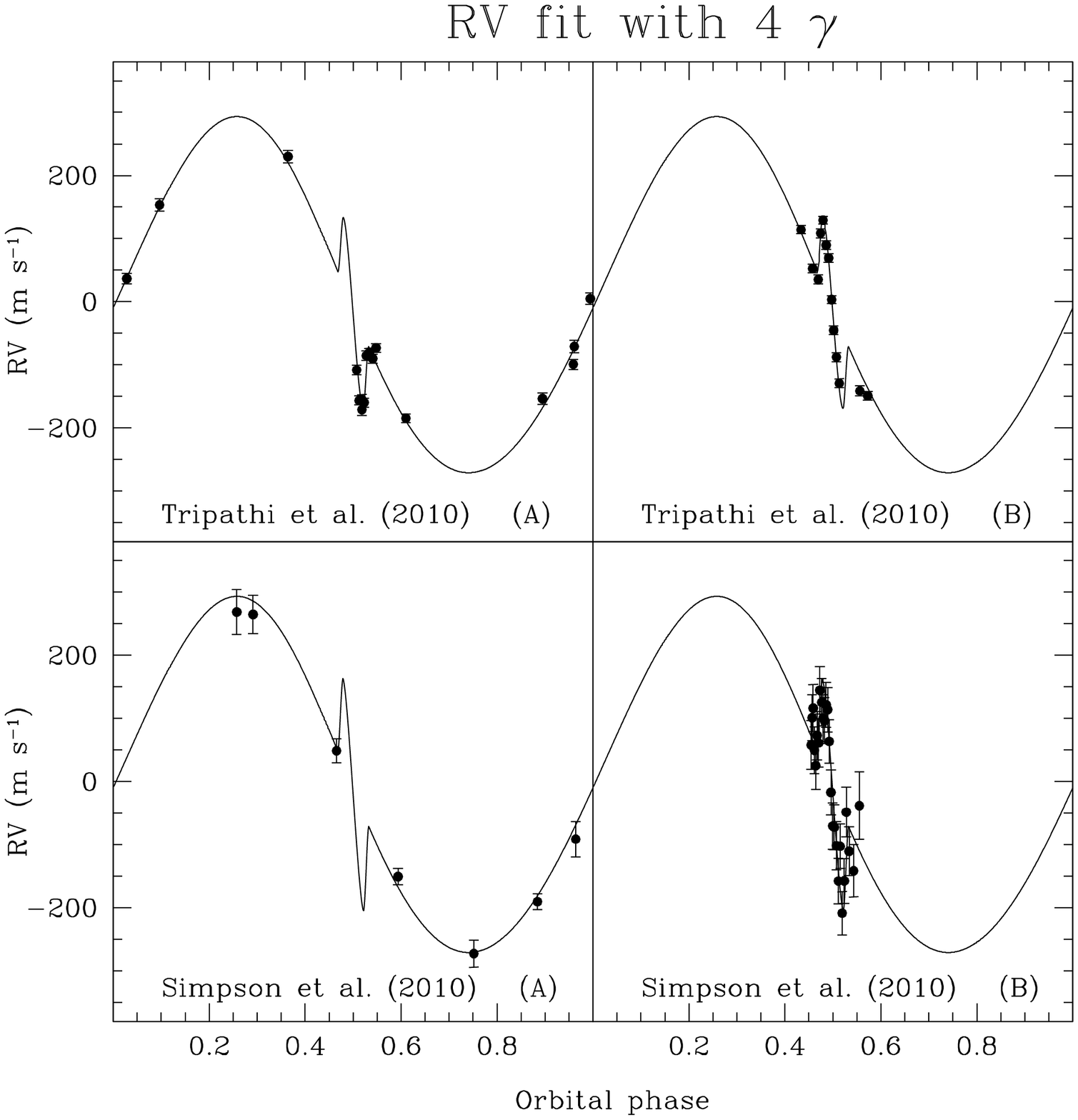}
\includegraphics[width=7.2cm]{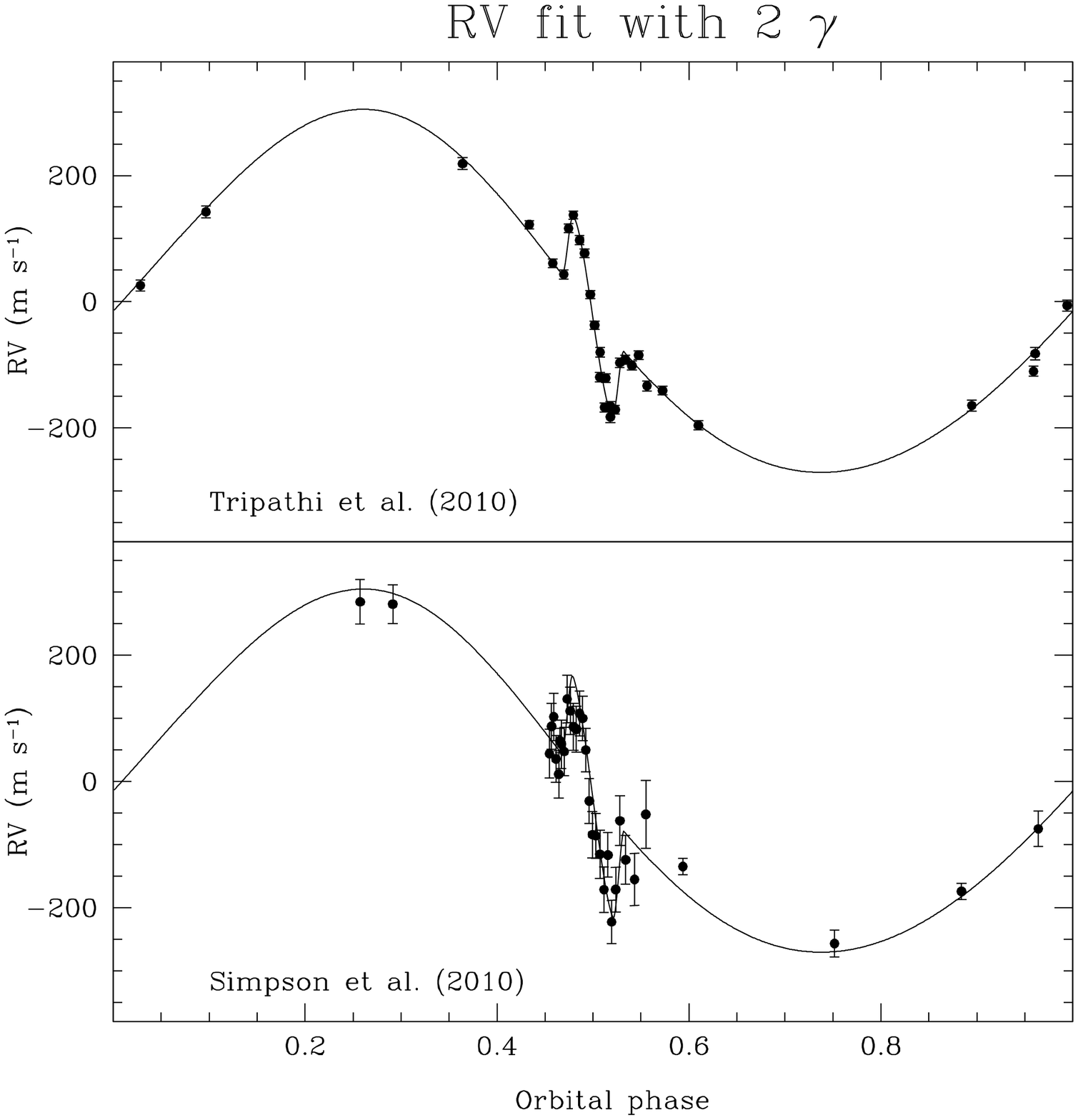}
\includegraphics[width=7.2cm]{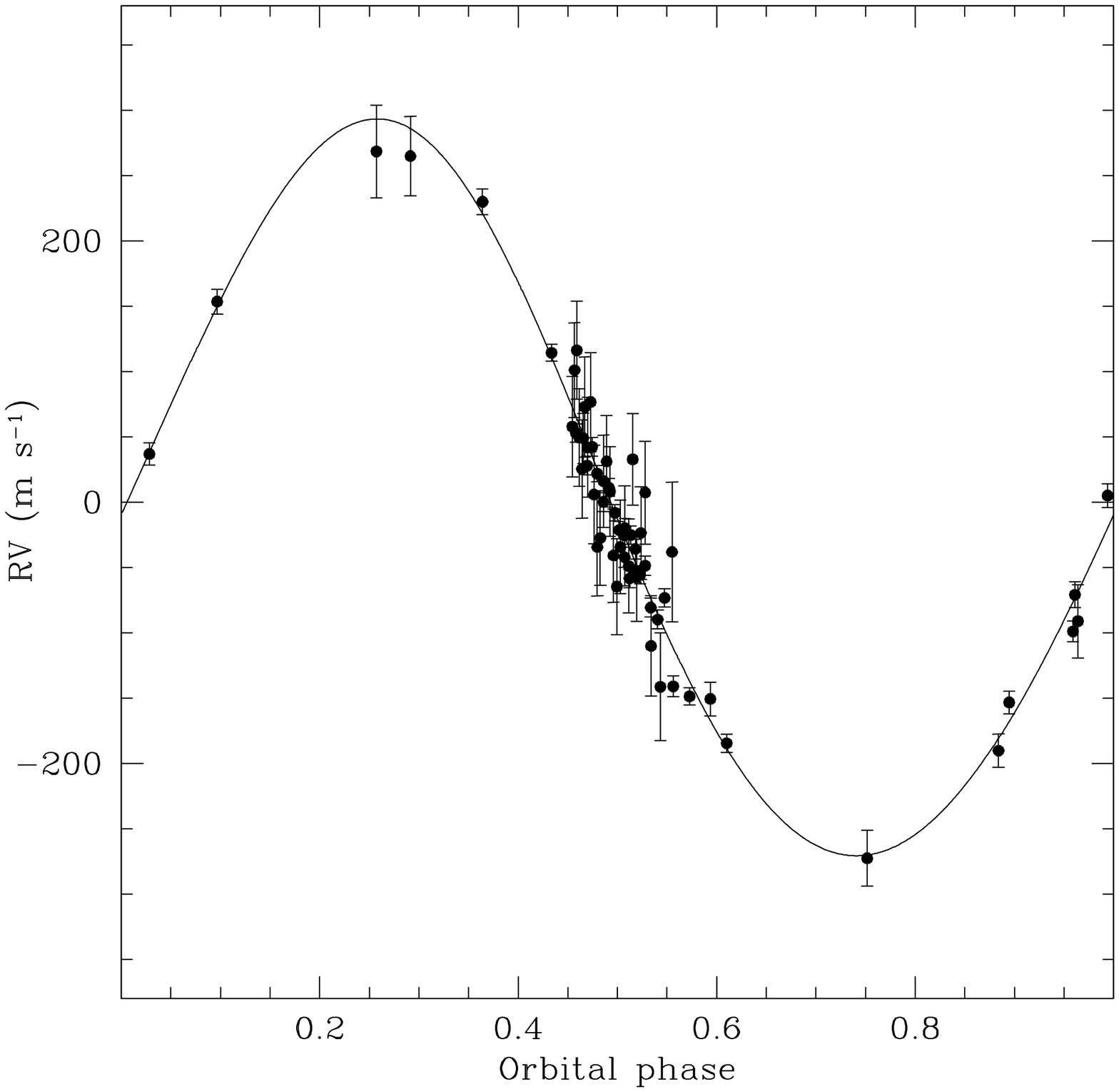}
\includegraphics[width=7.2cm]{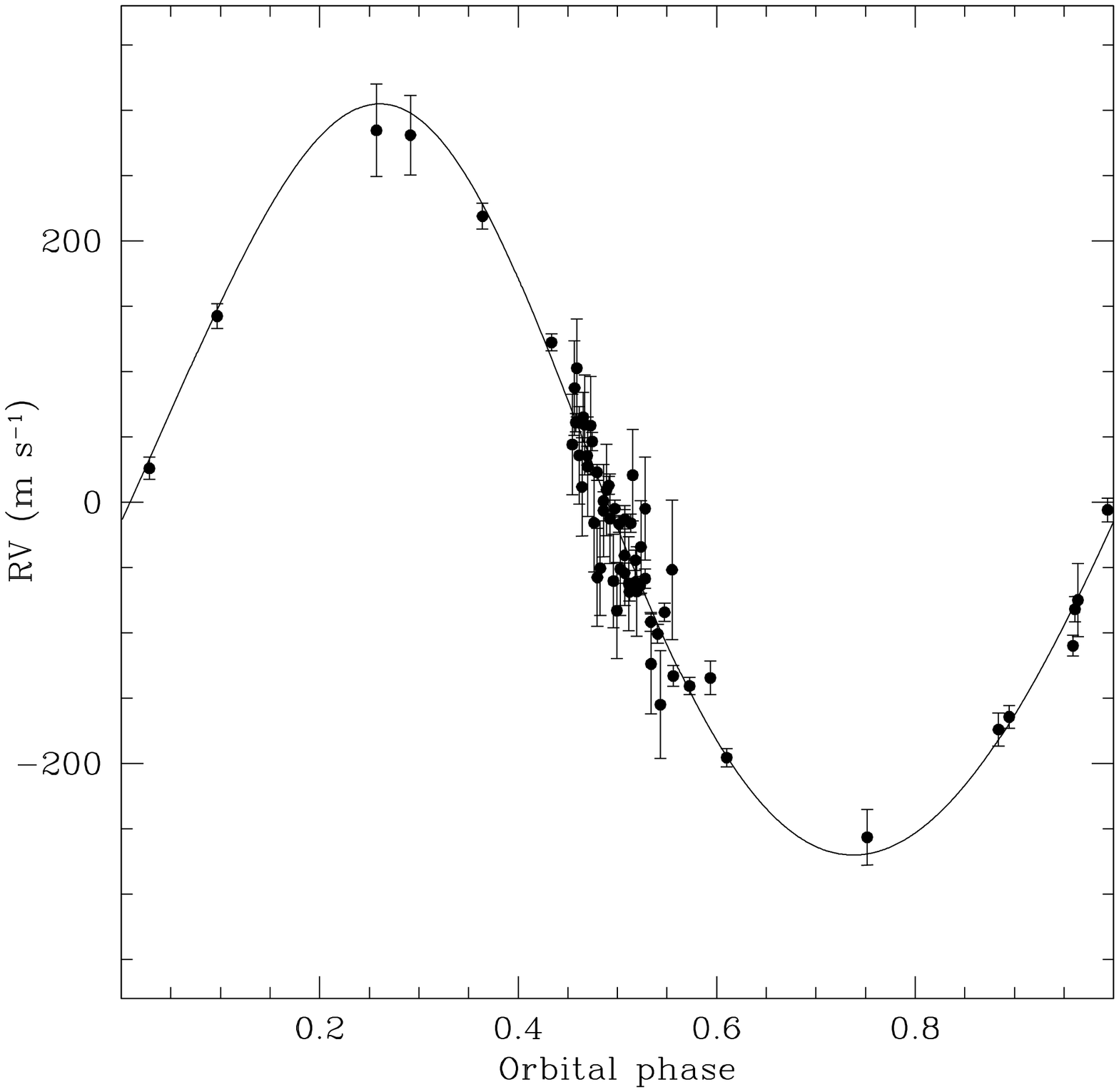}
\includegraphics[width=7.2cm]{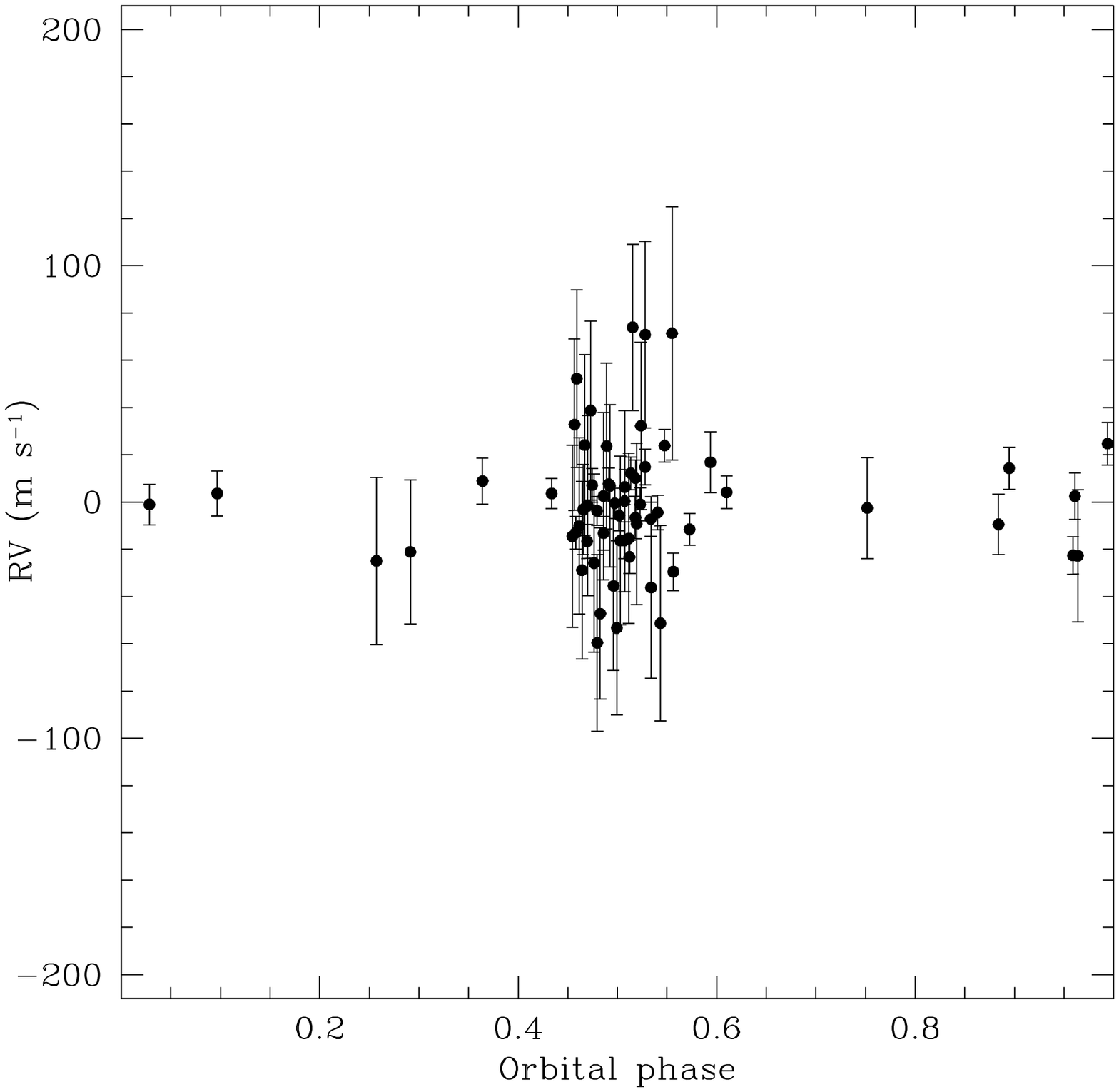}
\includegraphics[width=7.2cm]{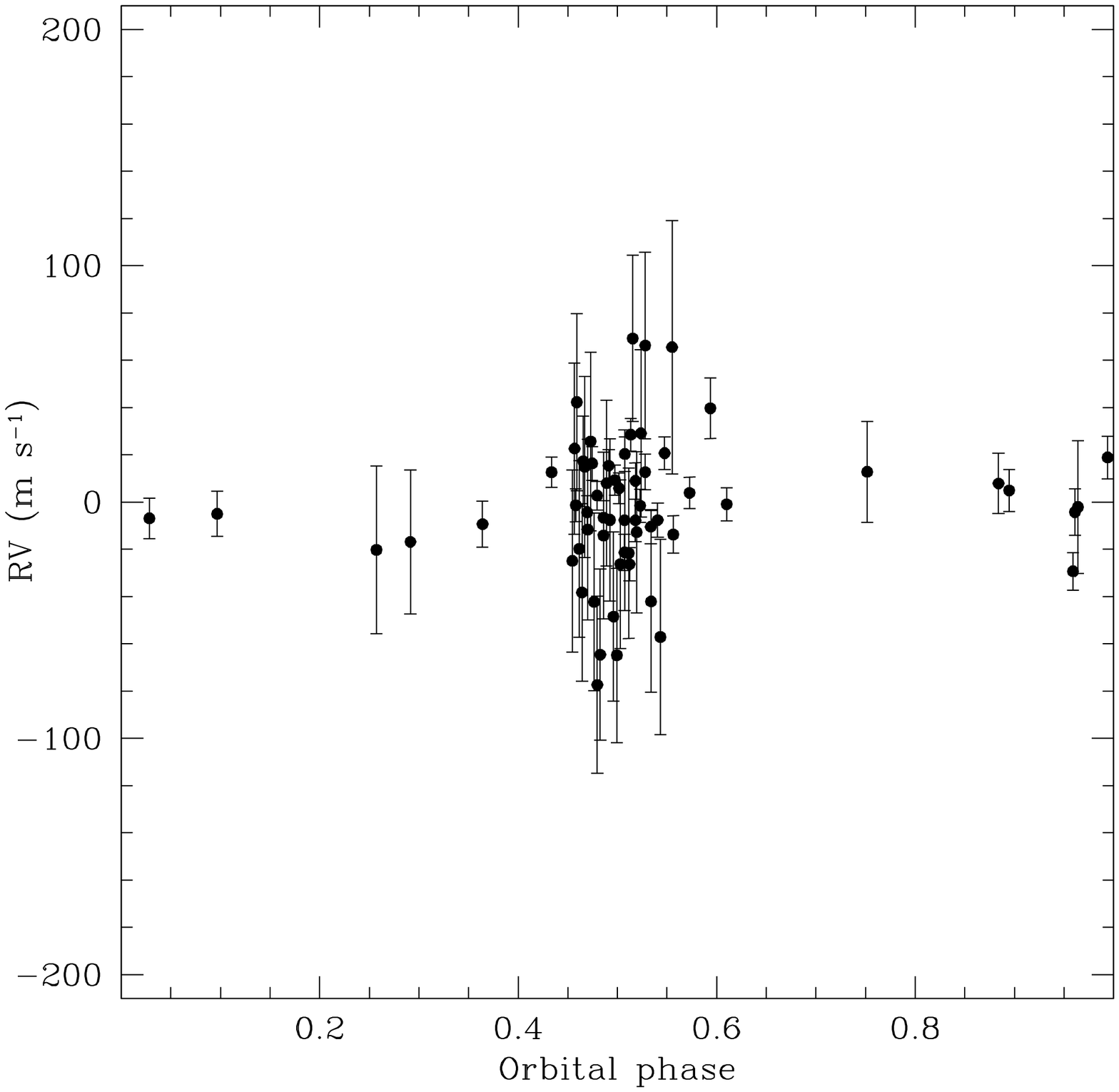}
\caption{
Radial velocity fit  once considering four and two  values of $\gamma$
as described  in the text  (left and right panels  respectively).  Top
panels:  radial   velocity  measurements  along  with   our  best  fit
models. Barycentric radial  velocities are subtracted.  Middle panels:
all radial velocity measurements  after subtraction of the barycentric
velocities and the  Rossiter-McLaughlin anomaly.  Lower panels: radial
velocity measurements after subtraction of the barycentric velocities,
the Rossiter-McLaughlin anomaly and the Keplerian orbit.
}
\label{fig:rv}
\end{center}
\end{figure*}

The convergence toward the best-fit  solution was obtained by means of
a Levenberg-Marquardt  algorithm, and  the uncertainties and  the best
fit values  of the parameters by  means of a Markov  Chain Monte Carlo
analysis  as  done for  the  photometric  data.   The radial  velocity
measurements  after subtraction of  the barycentric  velocities, along
with the best fit model  and distinguished in the four (two) different
groups of data to which  the different $\gamma$ were applied are shown
in Fig.~\ref{fig:rv} (upper panels).  The result after the subtraction
of the RM anomaly is  shown in the middle panels of Fig.~\ref{fig:rv},
and the residual velocities after subtracting the Keplerian orbit also
are shown in the bottom  panels.  Our best-fit parameters are given in
Table~\ref{tab:results_RV}  along with  the values  obtained  by other
authors.  Our best-fit  model corresponds to a $\sqrt{\chi_{r}^2}=1.5$
for the four $\gamma$  solution and to $\sqrt{\chi_{r}^2}=1.6$ for the
two  $\gamma$  solution.  Our   results  are  in  agreement  with  the
literature values. The barycentric velocities for the case of the four
$\gamma$  solution are consistent  with those  derived by  Tripathi et
al.~(2010) and Simpson et al.~(2010), and in the case of the 2$\gamma$
solution  our values  for  each  dataset are  in  between the  results
reported by those authors for their  own data.  We obtained a value of
the spin-orbit  angle consistent  with zero. We  also notice  that the
rotation  velocity we obtained  for the  four $\gamma$  solution ($\rm
v\sin (i)=13.9^{+0.3}_{-0.5}$) is perfectly consistent with the result
of      Miller     et      al.~(2010)     implying      $\rm     v\sin
(i)=13.9^{+0.03}_{-0.03}$. For  the two $\gamma$  solution we obtained
instead  a   larger  value  of  the  rotation   velocity  ($\rm  v\sin
(i)=14.5^{+0.3}_{-0.3}$).

\begin{table*}
\caption{
Best-fit  parameters  obtained  from  our  reanalysis  of  the  radial
velocity measurements (TW=This work), and from the study of Simpson et
al.~(2010, SI10),  Tripathi et al.~(2010, TR10),  Miller et al.~(2010,
MI10) and Pollacco et al.~(2008, PO08).
\label{tab:results_RV}
}
\begin{center}
\resizebox{17.5cm}{!}{
\begin{tabular}{c c c c c c c c c c}
\hline
\hline
        $v\sin i$    & $\lambda$ &  $\tilde{K}$ &  $k$  &  $h$  & $\gamma_1$ & $\gamma_2$ & $\gamma_3$ & $\gamma_4$ & Ref \\
        (km$\rm \,s^{-1}$) &     (deg) & (m$\rm \,s^{-1}$) &      &       &  (km$\rm \,s^{-1}$) & (km$\rm \,s^{-1}$) & (km$\rm \,s^{-1}$) & (km$\rm \,s^{-1}$) & \\
\hline
 13.9$^{+0.3}_{-0.5}$ & -3$^{+1}_{-2}$ & 282$^{-5}_{+7}$ &  0.04$^{+0.02}_{-0.01}$ & 0.03$^{+0.01}_{-0.01}$ & 0.029$^{+0.007}_{-0.002}$ & 0.048$^{+0.003}_{-0.007}$ & -5.453$^{+0.005}_{-0.010}$  & -5.483$^{+0.01}_{-0.007}$  & TW (4$\gamma$)\\
 & & & & & & & & \\ 
 14.5$^{+0.3}_{-0.3}$ & -1.9$^{+1.4}_{-0.9}$ & 287$^{+3}_{-9}$ &  0.060$^{+0.009}_{-0.024}$ & 0.035$^{+0.007}_{-0.016}$ & 0.040$^{+0.003}_{-0.004}$ & - & -5.469$^{+0.007}_{-0.005}$ & - & TW (2$\gamma$)\\
 & & & & & & & & \\ 
 15.7$^{+1.4}_{-1.3}$ & 13$^{+9}_{-7}$ & 276$\,\pm\,$11 & - & - & - & - & -5.458$\,\pm\,0.007$ & -5.487$\,\pm\,0.009$ & SI10 \\
 & & & & & & & & \\ 
 14.1$^{+1.5}_{-1.3}$ & 3.3$^{+2.5}_{-4.4}$ & 290.5$^{+9.8}_{-9.2}$ & - & - & 0.0335$^{+0.0063}_{-0.0045}$ & 0.0476$^{+0.0062}_{-0.0069}$ & - & - & TR10 \\
 & & & & & & & & \\ 
 13.9$^{+0.03}_{-0.03}$ & 5$^{+6}_{-5}$ & 278.2$^{+13.8}_{-13.4}$ & - & - & - & - & -5.4599$^{+0.0037}_{-0.0036}$ & - & MI10 \\
 & & & & & & & & \\ 
 13.4$\,\pm\,$1.5 & - & 251.2$^{+7.9}_{-10.8}$ & - & - & - & - & -5.4887$^{+0.0013}_{-0.0018}$ & - & PO08 \\
\hline
\end{tabular}
}
\end{center}
\end{table*}

%

%

\section{Analysis of the (O-C) transit timing diagram}
\label{s:oc}

\def\figPSD{
\begin{figure}
\includegraphics[width=0.48\linewidth,viewport=0 0 210 170,clip]{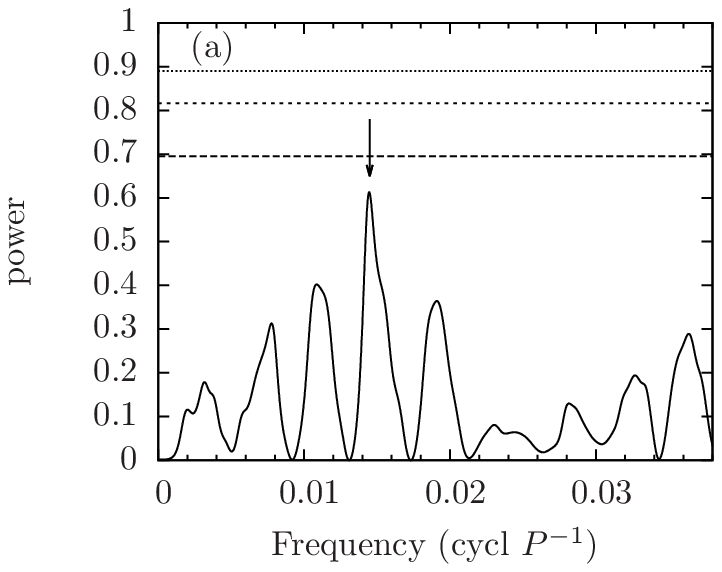}
\includegraphics[width=0.48\linewidth,viewport=0 0 210 170,clip]{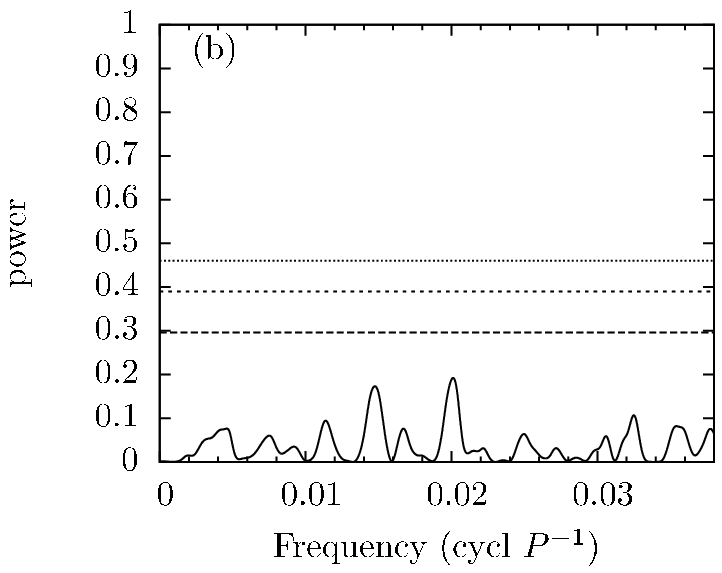}
\caption{\label{FigPSD} Periodograms of the TTV signal. (a) Considering
only the data used in Maciejewski  et al. 2010. (b) Using all the data
present  in  Tab.~\ref{tab:midtransits}.  The horizontal  lines  (from
bottom to top) give the FAP thresholds 0.1, $10^{-2}$, and $10^{-3}$.}
\end{figure}
}

\def\figMmax{
\begin{figure*}
\begin{center}
\includegraphics[width=0.45\linewidth,clip]{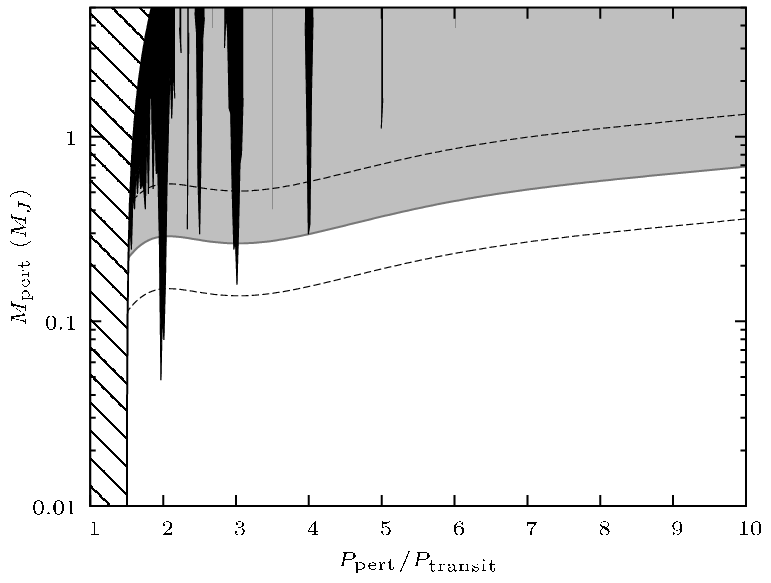}
\includegraphics[width=0.47\linewidth,clip]{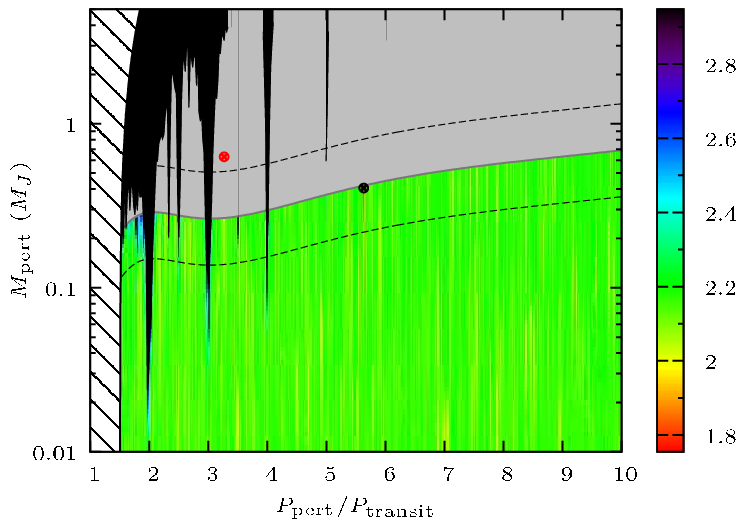}
\caption{
\label{FigMmax} Left: constraints on the maximal mass of a perturber:
99$\%$ confidence detection limits from periodograms of transit timing
(black) and radial velocity data (grey).  The dashed lines denotes the
uncertainty  range  of  the  radial velocity  detection  region.   The
hatched   lines    indicates   unstable   orbits.     Right:   reduced
$\sqrt{\chi^2_r}$ values  resulting from the  fit of the  observed O-C
timing residuals with our model.  The calculation is extended over the
radial velocity  undetectability region, but we also  show the regions
that  produce  a $3\,\sigma$  in  black.   Notice  that these  regions
overlap with the transit timing detectability regions of the
left panel.
}
\end{center}
\end{figure*}
}

In a first step, we analyse  whether or not a quadratic departure from
a linear fit is present in the transit timings. This could result from
the  direct  interaction  with   a  perturber  on  an  extended  orbit
(Borkovits et al.  2011), or  from the light travel timing produced by
the  motion  of  the  star  also induced  by  a  hypothetical  distant
companion (Montalto~2010).   This test can be  performed following the
approach of Pringle (1975) which  measures the improvement of a fit by
a quadratic parabola with respect to a simpler one by a straight line.
However, the quadratic coefficient obtained by least square is already
zero within  the errorbars $(-0.9 \pm 3.5)\,\times  10^{-9}$ days.  We
thus conclude that  there is no significant long  term quadratic trend
in the data.

Then, we follow the analysis of Maciejewski et al. (2010).  We compute
a  Lomb-Scargle  periodogram on  the  Transit  Timing Variation  (TTV)
signal in  order to detect  a periodic oscillation that  would reflect
the perturbation  of a  close-in undetected body  in the  system.  For
that purpose, we use the generalized version (GLS) of the Lomb-Scargle
periodogram (Zechmeister \&  K$\rm \ddot{u}$rster 2009). Basically the
GLS fits a sinusoid to the  data for each frequency by using the least
square method, as the Lomb-Scargle  algorithm, but in addition to that
it  allows   for  the  presence   of  an  additional   constant  term.
False-alarm  probabilities  (FAP)   are  estimated  by  computing  GLS
periodograms on a  large number of sets of  artificial observations in
which, for  each epoch where a  transit has been  observed, we replace
the measured (O-C) by a  random value normally distributed around zero
with a standard deviation equal to the uncertainty of that point.  The
number of periodograms  containing a peak with an  power above a given
threshold out of the total  number of trials represents our estimation
of the FAP for that given threshold.

Fig.~\ref{FigPSD}(a)   shows  the   GLS   periodogram  obtained   when
considering  only the  transits  used by  Maciejewski  et al.   (2010,
Fig.~3).   We   obtain  a  dominant   peak  at  a   frequency  $f_{\rm
  TTV}=0.0145$   cycle   $P^{-1}$,    which   corresponds   to   P$\rm
_{TTV}\,=\,127$ days and  an power of 0.61, as shown  by the arrow, in
complete  agreement with  the  result of  Maciejewski  et al.  (2010).
Nevertheless the  false-alarm probability associated to  that power is
27$\%$.  There is thus more than one  chance out of 4 for this peak to
be  fortuitous.   For  the   sake  of  completeness,  the  false-alarm
probability   thresholds  of   0.1,  $10^{-2}$,   and   $10^{-3}$  are
represented  by   three  horizontal  lines   in  the  two   panels  of
Fig.~\ref{FigPSD}. We did again the same analysis with all the data of
the  table~\ref{tab:midtransits}.    The  results  are   displayed  in
Fig.~\ref{FigPSD}(b).  In  that case, the peak with  the highest power
is  now at $f_{\rm  TTV}=0.0201$ cycle  $P^{-1}$ with  a FAP  equal to
56$\%$.   It thus  seems  that the  TTV  signal does  not contain  any
significant periodic oscillations.

\figPSD

\section{Analysis of the (O-C) radial velocity diagram}
\label{s:RVres}

We  initially  checked  for the  presence  of  either  a linear  or  a
quadratic term  in the  (O-C) radial velocity  residuals by  using the
Pringle  (1975)  test.   We  considered  the  2$\gamma$  solution  and
obtained that in both cases  the coefficients are consistent with zero
($0.00001\,\pm\,0.00012$     for    the     quadratic     term)    and
($-0.030\,\pm\,0.035$ for the linear term).

\begin{figure}
\begin{center}
\includegraphics[width=0.37\linewidth,angle=270]{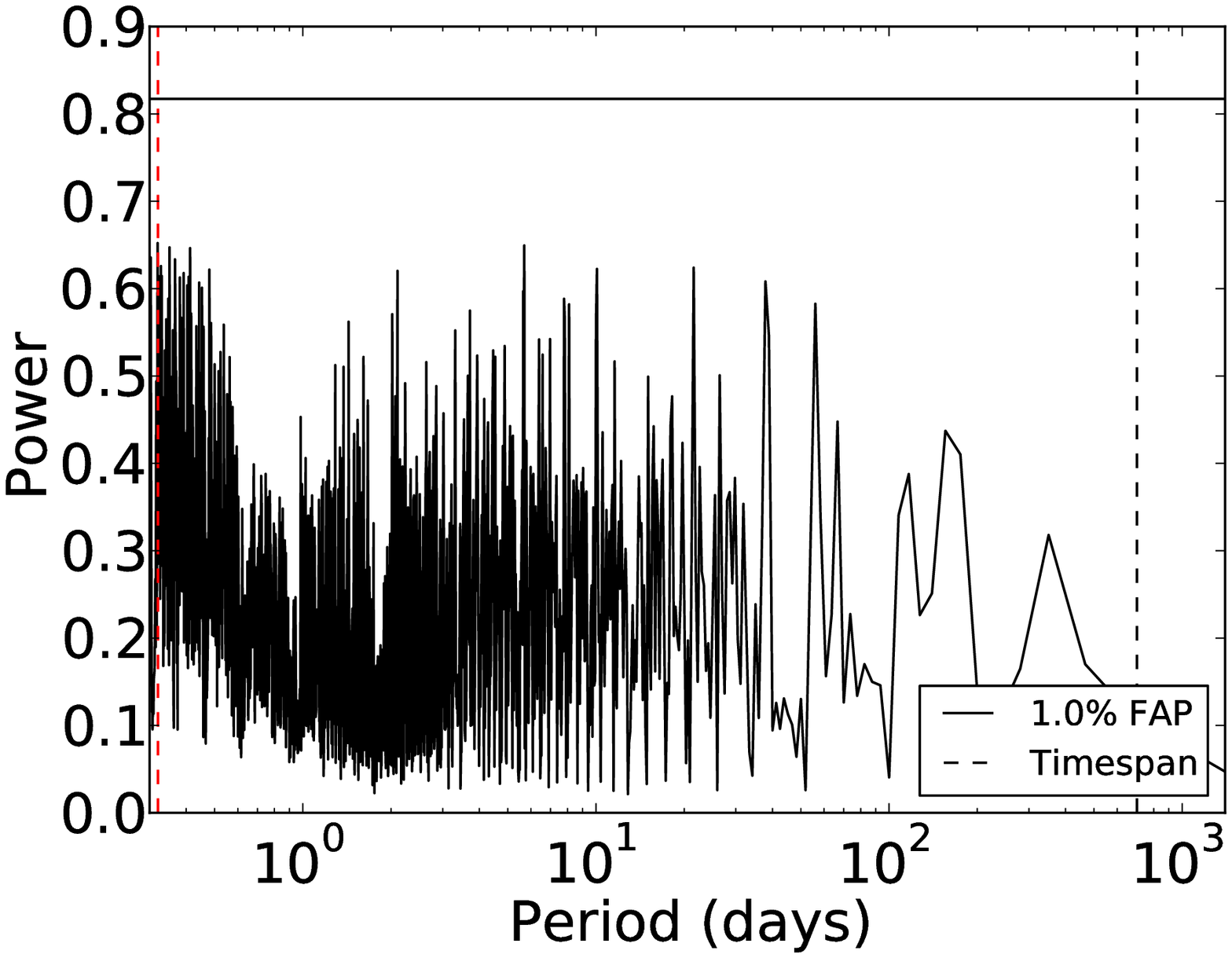}
\includegraphics[width=0.37\linewidth,angle=270]{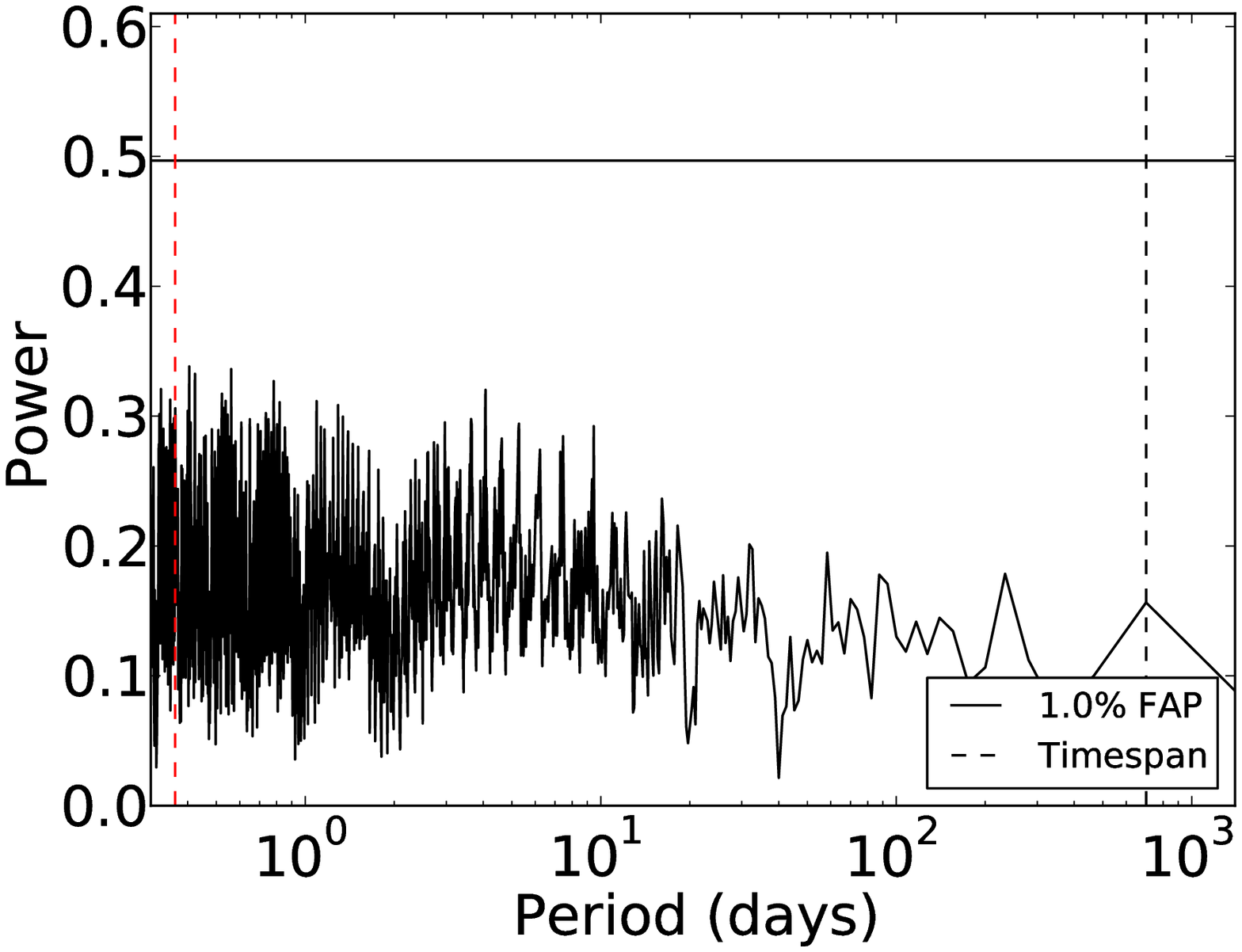}
\caption{\label{fig:RVresper}  Periodogram   of  the  radial  velocity
residuals  considering   the  4$\gamma$  solution   (left)  and  the
2$\gamma$ solution  (right). The  continuous line denotes  the 1$\%$
FAP while the dashed line the time-span of the observations.}
\end{center}
\end{figure}

A GLS periodogram was then computed.  We considered initially the case
of   the   residuals   obtained   fitting   the   4$\gamma$   solution
(Sect.~\ref{s:analysis_RV}).   As seen in  Fig.~\ref{fig:RVresper}, no
significant peaks can  be found in the periodogram.   The highest peak
is at 0.35 days and has a FAP of 31$\%$.  Alternative, considering the
residuals obtained by the  2$\gamma$ solution, we obtained the highest
peak  at 0.36  days  with a  FAP=39$\%$.   The FAP  of  the peaks  are
estimated  with  a bootstraping  method  in the  same  way  as in  the
previous section. The only difference  is that the artificial data are
made by  shuffling (with repetition) the residuals  instead of drawing
random values from a normal distribution.


\section{Discussion}
\label{s:discussion}

Since the  TTV signal does not  present any long  term variations, nor
short period oscillations, one  cannot assert that the system actually
contains an  additional planet. And,  if such a companion  does exist,
its orbital parameters and its  mass are poorly constrained due to the
lack   of   expected  patterns   in   the   current  available   data.
Nevertheless,   the   residuals   of   the   fit   are   quite   large
($\sqrt{\chi_r^2}  \simeq 2.30$ and  $\sqrt{\chi_r^2} \simeq  1.5$ for
transit and RV respectively).   As noticed also by other investigators
in the  past these values  do not indicate  a satisfactory fit  to the
(O-C).  The model is thus not  complete.  Several causes can be at the
origin of this  result among which we consider  here stellar activity,
the  presence of an  additional planet  or exomoon  and underestimated
timing uncertainties.

\subsection{Stellar activity}
A possible source  of TTVs can be the activity  of WASP-3. Indeed, the
existence of  spots on the surface  of the star,  partially covered by
the  planet  during  transits,  should  produce  fluctuations  in  the
luminosity leading to some errors in the determination of the times of
transit  minimum  (e.   g.    Sanchis-Ojeda  et  al.~2011;  Oshagh  et
al. 2012).  Moreover, the spots, if they exist, should not be the same
between the beginning  and the end of the  observations given the long
time  span that  has  been  covered ($\sim  4.5$  years).  This  would
explain  why  no  periodic   oscillation  is  detected.   Tripathi  et
al.~(2010) reported  fractional transit depth variations  of the order
of $7\%$,  even if the same  authors were not  confident whether these
variations  were genuine  or due  to  systematics in  their data.   In
addition, they report a  mean logR$^{'}_{\rm HK}$=-4.9 from their Keck
spectra taken in 2008-2009.

On the other  hand, we reanalysed the spectra  taken with SOPHIE. From
the 2007  observations (Pollacco et  al.~2008) we derived  a $\log{\rm
  R}^{'}_{\rm HK}$ value of  -4.95, whereas the 2009-2010 observations
(Simpson et al.  2010) provided  a higher value for the activity index
of $\log{\rm  R}^{'}_{\rm HK}$=-4.80.   Therefore it appears  that the
mean activity level of the star changed during these years approaching
an active phase in 2010.  Once considering also the upper limit on the
stellar age and  the rotation period reported by  Miller et al.~(2010,
age$<2$  Gyr and  $\rm  P_{rot}=$  4.3 days)  the  presence of  active
regions on this  star may not appear a  rare circumstance.  Despite no
clear  evidence of  starspots  crossing has  been  reported yet,  this
analysis  clearly  claims  for  a  more intensive  monitoring  of  the
activity  level  of  WASP-3  in  order to  understand  its  impact  on
photometric and radial velocity measurements.

\subsection{Additional planet}
\figMmax

We now give  some constraints on the mass  of a hypothetical planetary
perturber.  For  that, we  use both the  dispersion of the  O-C radial
velocity  and  photometric diagrams.   For  what  concerns the  radial
velocity  residuals  we  adopted  here  the results  coming  from  the
2$\gamma$ solution, since no significant difference was found adopting
instead the 4$\gamma$ solution.

\subsubsection{Radial velocities}

In the literature, two main  approaches are used to find the detection
limits  in  radial velocity  data.   One  is  based on  $\chi^2$-  and
$F$-tests (e.g.  Lagrange et al.  2009, Sozzetti et al. 2009), another
is based  on a  periodogram analysis (Cumming,  Marcy \&  Butler 1999,
Endl et al. 2001, Cumming  2004, Narayan, Cumming \& Lin 2005).  Here,
the second approach was chosen due to the number of measurements which
is considered high enough for a reliable periodogram analysis.

For each period, a fake  eccentric planetary signal is inserted in the
data, while the original data is treated as random noise. On these new
RV  series,  the  power   (in  the  periodogram)  is  calculated.  The
semi-amplitude of the  fake signal is changed untill  the FAP level is
reached  for all  eccentricities $e$,  times of  periastron  $T_c$ and
longitudes  of periastron  $\varpi$. In  this paper,  a FAP  of $1\%$,
determined with $1000$ shuffled time series, is used. Fake signals are
tested for periods $P$ between 1  and 20 days. The orbital elements of
the eccentric signals range, in 10 steps, as follows: $0\leq e\leq 1$,
$0\leq  T_c  \leq  P$  and  $0  \leq  \varpi  \leq  2\pi$.  The  final
semi-amplitude can be transformed  in planetary mass and expresses the
lower limit for detectable planets at that period with these data.
\begin{equation}
M_p\sin i = 1.2\cdot 10^{-3} K\sqrt{1-e^2}\left( \frac{P M_{\ast}^2}{2\pi G}\right)^{1/3}
\end{equation}
%
with the planetary mass in  Earth mass, the semi-amplitude $K$ in m/s,
the period  $P$ in days, the  stellar mass $M_{\ast}$  in solar masses
and   the  gravitational   constant  $G$   in  m$^3$kg$^{-1}$s$^{-2}$.
Therefore the  continuous grey line in  Fig.~\ref{FigMmax} denotes the
limit  in the perturber  mass beyond  which a  signal would  have been
detected in the radial velocity  data with a confidence limit equal to
$99\%$. The dashed lines shows the 1-$\sigma$ uncertainty range of the
radial velocity detection limit.

\subsubsection{Photometry}

We  exclude compact systems  which lead  to unstable  evolutions. Only
circular and coplanar systems  have been considered since they provide
the strongest  constraints and because the  projected spin-orbit angle
measure  on WASP-3b  by the  Rossiter-McLaughlin effect  is compatible
with  zero  as demonstrated  above,  suggesting  that  if a  planetary
companion exists, the system is likely coplanar.

The radial  velocity measurements are  used to exclude  any perturbers
that would  induce RV signal with  an amplitude larger  than the 1$\%$
FAP threshold.  The same exercise has  been performed with  the O-C of
the transit timing measurements. For that, we simulated a large number
of O-C  on a grid of  parameters of the perturber.   We considered 400
periods $P_{\rm  pert}$ ranging between  1.5 and 10  $P_{\rm transit}$
(where $P_{\rm  transit}=P$ is the  period of the  transiting planet),
and 100 masses $M_{\rm  pert}$ evenly distributed in logarithm between
0.01  and 5  $M_J$.   For each  period  and mass,  60 simulations  are
performed   with   different   initial   longitudes  between   0   and
360$^{\circ}$.   Among   the  60  simulations,  the   one  giving  the
periodogram  with  the lowest  maximum  amplitude  is  kept.  If  this
amplitude  is above  the  1$\%$ FAP  threshold  determined in  section
\ref{s:oc}, then the corresponding perturber should have been detected
in the  periodogram of the O-C (see  Fig.~\ref{FigPSD}), otherwise the
perturber can exist but it is not detectable.

Figure~\ref{FigMmax}  (left) shows the  results.  The  hatched region,
which  extends up to  a period  ratio of  1.5, delineates  the chaotic
orbits  which are  excluded.  The  boundary  of this  region has  been
derived from the stability criterion of (Gladman 1993). The gray curve
fixes  the limit  of the  perturber's  mass from  the radial  velocity
measurements. Any perturber in the gray region would induce a periodic
RV signal with a significant amplitude. And finally, the black regions
delineate the perturbers that produce TTVs with significant oscillating
terms.  Combining  all the information,  it turns out that  the radial
velocity technique excludes any perturber  as massive as Jupiter up to
a  period ratio  of  10,  and the  TTV  measurements provide  stronger
constraints close to mean motion resonances.

In a  second step, we  focused on perturbers  that do not  produce any
significant  periodic  RV  signals  nor periodic  TTV  signals.   Such
perturbers  are  located in  the  white  region of  Fig.~\ref{FigMmax}
(left). For those perturbers, we  check whether they can reproduce the
O-C transit timing  diagram or not.  In that purpose,  we use the same
grid  of parameters  as  in Fig.~\ref{FigMmax}  (left),  and for  each
initial  conditions, we  compute  the expected  TTVs  and the  reduced
chi-square  with  respect  to   the  observations.   The  results  are
displayed  in Fig.~\ref{FigMmax}  (right).   The hached  and the  gray
regions  are the  same  as in  Fig.~\ref{FigMmax}  (left).  The  black
regions correspond  now to simulated TTVs above  the 3-$\sigma$ level.
This  threshold is  obtained  from the  $\chi^2$-distribution with  33
degrees   of  freedom.    It   corresponds  to   $\chi_r^2=\chi_{r,\rm
  min}^2(1+1.821)$,  or   $\sqrt{\chi_r^2}=2.95$.   The  colour  scale
represents the $\sqrt{\chi_r^2}$  from the lowest values in  red up to
the 3-$\sigma$  threshold in  dark violet.  The  red circle  shows the
best  fit to  the observation  with  $\sqrt{\chi_r^2}=1.76$.  However,
such a perturber, with a mass  of 0.63 $M_J$ should have been detected
in the radial velocity analysis.  The best fit within the undetectable
perturbers  is just  below  the RV  detection  threshold with  $M_{\rm
  pert}=0.41  M_J$ and  $P_{\rm pert}=5.63  P_{\rm transit}$,  but the
corresponding reduced  chi-square is only  $\sqrt{\chi_r^2}=1.83$. The
improvement is  very weak. Moreover,  from Fig.~\ref{FigMmax} (right),
one can see that such values of the reduced chi-square are spread more
or less randomly within the undetectable region.

As  noted by  Maciejewski et  al.  (2010),  the presence  of  an outer
companion less massive than WASP-3b  but still on a short period orbit
would   make  the  system   quite  unusual.    Multiplanetary  systems
containing at least  a Jupiter-mass planet are indeed  much wider, and
the  less  massive   planet  is  usually  the  closest   to  the  star
(e. g. Lissauer et al. 2011).

\subsection{Exomoon}
An exomoon is also supposed  to generate a periodic oscillation in the
TTV. However,  Maciejewski et al.  (2010) have already  discarded this
hypothesis  since the  transits do  not show  any  duration variations
shifted   in   phase  by   $\pi/2$   with   respect   to  the   timing
variations. Here,  we perform a  more detailled analysis based  on the
results of Kipping~(2009).

First of all, we check that  an exomoon can have a stable orbit. 
If the moon is less than twice as dense as the planet the minimum
distance of the moon is set by the Roche limit. 
Let $\xi$ be the semi-major axis of a
hypothetical satellite  divided by the  Hill Radius, i.e. $a_s  = \xi
R_H$ where  $a_s$ is the  semi-major axis of  the moon and  $R_H$  the
Hill Radius.  According to  Kipping~(2009), $\xi$ should  satisfy the
following inequality
\begin{equation}
\chi_{\rm min} \lesssim \xi \lesssim \frac{1}{3}\ ,
\end{equation}
where $\xi_{\rm min}  = 1/186 \times(M_s/M_\oplus)^{-0.063} (P/1 {\rm
  d})^{-2/3}$ represents the Roche limit. In this expression, $M_s$ is
the mass of the satellite and $P$ is the orbital period of the planet.
For an exomoon of the mass of the Earth's Moon, we get $\xi_{\rm min}
= 0.0047$,  and for an  exomoon of the  mass of the  Earth, $\xi_{\rm
  min}  = 0.0036$.   In  both  cases $\chi_{\rm  min}$  is lower  than
$1/3$. 
For moons which density is more than twice that
of the planet the Roche limit would be inside the planet, therefore
the minimum distance would correspond to the planetary radius and the
above inequality would be automatically satisfied.
Therefore an  exomoon can exist on a stable  orbit around WASP-3.

Then, we  estimate the maximal  RMS amplitude ($\delta_{\rm  TTV}$) of
TTV   that   an   exomoon   on   a   coplanar   circular   orbit   can
produce. According to Kipping (2009), this amplitude is given by
\begin{equation}
\delta_{\rm TTV} = \frac{1}{\sqrt{2}} \frac{P}{2\pi} 
\left(\frac{M_{ps}}{3 M_\star}\right)^{1/3} \mu(1-\mu)^{1/3} \xi\ ,
\end{equation}
with $\mu = M_s/M_{ps}$ and $M_{ps}$ is the sum of the planet mass and
the satellite  mass. Without any  constraint on $\mu$, the  maximum of
the product $\mu(1-\mu)^{1/3}$ is attained for $\mu=3/4$, and is equal
to $3/4^{1/3}$. However,  by definition, the Moon should  have a lower
mass than  the planet.   If the transit  lightcurves are those  of the
planet,  $\mu$  should  be   lower  than  $1/2$,  and  probably  much
lower. But let  us assume that $\mu=1/2$, this  will provide the upper
limit  of  $\delta_{\rm TTV}$.  If  the  planet  has an  exomoon,  the
Keplerian  orbit  derived in  the  previous  section  is that  of  the
planet-satellite  barycenter around  the star.   Thus the  fitted mass
corresponds to  $M_{sp}$. Using $M_{sp}/M_\star  = 1.5\times 10^{-3}$,
one obtains
\begin{equation}
\delta_{\rm TTV} \leq \frac{1}{\sqrt{2}}\frac{1}{2^{4/3}}
\frac{P}{2\pi} \left(\frac{M_{sp}}{3M_\star}\right)^{1/3} \xi
 = 9.4 \,\xi \quad {\rm [min]}\ .
\end{equation}
For $\xi = \xi_{\rm max} =  1/3$, this leads to $\delta_{\rm TTV} \leq
3.1$ min.  The result is larger than the observed RMS of the O-C which
is equal  to 1.1 min. Thus,  a ``satellite'' as massive  as the planet
($\mu=1/2$)  is able  to  produce significant  TTV  with an  amplitude
comparable to the observed one.

\begin{figure}
\begin{center}
\includegraphics[width=7.5cm]{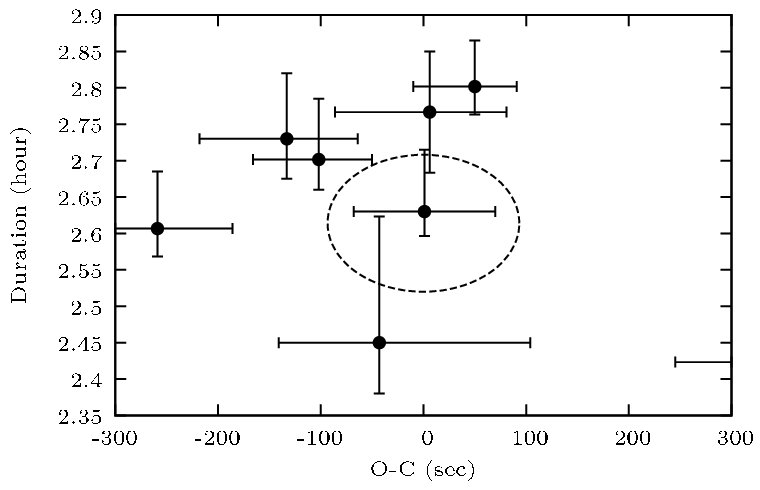}
\caption{
Transit  durations against O-C  transit timing  residuals for  the new
transits  presented in  this  work. Overplot  a representative  signal
produced by an exomoon having $M_s=0.35 M_J$ and period $\chi=1/3$.
}
\label{fig:duration_OC}
\end{center}
\end{figure}

We now  assume that the  observed O-C is  only due to  an hypothetical
satellite. Expecting  that this satellite  should have a  much smaller
mass than the  planet, we derive its mass  for $\xi=\xi_{\rm max}=1/3$
such that  $\delta_{\rm TTV} =  1.1$ min.  One gets  $\mu=0.14$, which
corresponds  to a mass  ratio of  $M_s/M_p=0.17$, or  $M_s\approx 0.35
M_J$. Thus, the lowest massive  satellite, on circular orbit, that can
account for the observed RMS of  the O-C is still large.  Assuming the
same  density as  a  giant planet  like  Jupiter, the  radius of  this
satellite should be  $0.46 R_J$. Such a big  satellite, if it existed,
would produce detectable transits  in the lightcurve.  We also noticed
that a  search for additional  transiting objects in the  NASA $EPOXI$
mission  WASP-3 lightcurve resulted  in a  null detection  (Ballard et
al.~2011). However, we note that accordingly to Domingos et al. (2006)
the value of the critical  semi-major axis could approach the value of
0.9309 in units of the Hill radius, for retrogade moons.  In this case
we can obtain  a more stringent constraint on the  maximum mass of the
moon which could be equal to $M_s/M_p=0.04$ or $M_s\approx 0.07\,M_J$.
In  Fig.~\ref{fig:duration_OC},  we also  show  our transit  durations
against the  O-C residuals.  In  case of an exomoon  being responsible
for the  claimed TTVs, we should  expect the observations  to trace an
ellipse in this diagram since  the TTVs and the TDVs (Transit Duration
Variations) produced  by an  exomoon are shifted  in phase  by $\pi/2$
(Kipping  2009).  For  illustration, we  overplot the  expected signal
that  the  above-mentioned  prograde  satellite  ($M_s=0.35  M_J$  and
$\chi=1/3$) should  generate.  Evidently given the  large errorbars of
our measurements  it is not  possible to explore this  possiblity, and
additional more accurate measurements are required for this analysis.

\subsection{Underestimated uncertainties}
It is difficult to ascertain  up to which level different instruments,
observing  conditions, reduction  and transit  fitting  procedures may
affect  the  results   reported  in  Table~\ref{tab:midtransits}.   To
address this point  it would be necessary to  homogeneously reduce and
analyze all the data collected so  far by all the different groups, an
approach which is not easy to put into practice.  In principle all the
transits considered here were  presented in referred journals and this
ensures that  accurate procedures like  those ones reported  here have
been applied  to estimate  transit timing errors.   It is  our opinion
however  that error  underestimation cannot  be completely  ruled out.
New observations will be certainly welcome to clarify this problem.

\section{Conclusions}
\label{s:conclusions}

In  this work we  provided a  throughout analysis  on the  presence of
additional  bodies  in the  WASP-3  system.  This  analysis serves  to
improve our  understanding of close-in  Jupiters and in  particular to
clarify if these planets are indeed isolated or not.

In addition to  present eight new transits of  WASP-3b acquired at the
Crow-Observatory-Portalegre   in  Portugal   we  reanalized   all  the
photometric and radial velcocity  measurements aquired so far for this
system.   We  concluded  that  there  is  no  convincing  evidence  of
additional  planetary  companions in  this  system;  both the  transit
timing and  the radial velocity  residuals do not  present significant
periodicities  (FAP=56$\%$  and  FAP=$31\%$  for  transit  and  radial
velocity in the best case scenario respectively) nor long term trends.

Combining  all transit  timing  and radial  velocity informations,  we
obtained that any  perturber more massive than $\rm  M \gtrsim 100 \rm
M_{\oplus}$ and  with period up  to ten times  the period of  the inner
planets is excluded at 99$\%$ confidence limit.

We  also investigated  the possible  presence  of an  exomoon on  this
system and determined that considering  the scatter of the O-C transit
timing residuals  a coplanar  exomoon would likely  produce detectable
transits,  an  hypothesis  that  can  be  ruled  out  by  observations
conducted by other researchers.  In case  the orbit of the moon is not
complanar the current accuracy  of transit timing and transit duration
measurements  prevents   to  make  any   significant  statement.   For
retrogade moons  the maximum mass  allowed at the  critical semi-major
axis is around 0.1 $M_J$.

Finally on  the basis of our  reanalysis of SOPHIE data  we noted that
WASP-3 passed  from a less active (  $\log{\rm R}^{'}_{\rm hk}=-4.95$)
to a more active ( $\log{\rm R}^{'}_{\rm hk}=-4.8$) state between 2007
and 2010.  Despite no clear spot  crossing has been  reported for this
system so far  we therefore pointed out the need  for a more intensive
monitoring of the  activity level of this star  in order to understand
its impact on photometric and radial velocity measurements.

Our lightcurves are made available through the on-line version of this
journal.

\section*{Acknowledgments}
This  work was  supported  by the  European Research  Council/European
Community  under  the  FP7  through Starting  Grant  agreement  number
239953,  and  through  grant reference  PTDC/CTE-AST/098528/2008  from
Funda\c{c}\~ao para  a Ci\^encia e  a Tecnologia (FCT,  Portugal). MM,
NCS  and SS  also acknowledge  the  support from  FCT through  program
Ci\^encia\,2007 funded  by FCT/MCTES (Portugal) and  POPH/FSE (EC) and
in   the  form  of   fellowship  references   SFRH/BDP/71230/2010  and
SFRH/BPD/47611/2008.   G.B.   thanks Paris  Observatory  and CAUP  for
providing  the necessary  computational resources  for this  work.  We
thank the  referee, Dr. David Kipping, whose  valuable comments helped
to improve this manuscript.

\section*{Appendix A: the normalized planet distance}

Here  we derive  the  expression for  the normalized planet  distance $z$ 
presented in Eq.~1. 
Assuming a circular  orbit and a constant projected  velocity ($v$) of
the transiting  planet on the plane  of the sky, in  any given instant
$t$ during the transit the normalized distance $z$ can be written as

\begin{eqnarray*}
z^2\, & = & \,\Big(\frac{b}{R_{\star}}\Big)^2\,+\,\Big(v\,\frac{t-T_0}{R_{\star}}\Big)^2, \nonumber \\
\end{eqnarray*}

\noindent
where $b$  is the impact parameter,  $R_{\star}$ is the  radius of the
star and  $T_0$ is the  time of transit  minimum.  At the time  of the
first or the fourth contact we have

\begin{eqnarray*}
z^2\,=\,(r+1)^2 & = & \Big(\frac{b}{R_{\star}}\Big)^2\,+\,\Big(v\,\frac{T_d}{2\,R_{\star}}\Big)^2, \\
\end{eqnarray*}

\noindent
where $r$ is  the ratio of the  radius of the planet to  the radius of
the star, and  $T_d$ is the total transit duration  (from the first to
the  fourth  contact).  Assuming  the projected  velocity  during  the
transit to be identical to the orbital velocity we can write

\begin{eqnarray*}
v & = & \sqrt{\frac{G\,M_{\star}}{a}}, \\
\end{eqnarray*}

\noindent
where $G$  is the gravitational  constant, $M_{\star}$ is the  mass of
the star, $a$ is the semi-major  axis and we neglected the mass of the
planet.  Eliminating in the  first equation above the impact parameter
derived from  the second, using  the third Kepler law  and introducing
the definition of the mean stellar density

\begin{eqnarray*}
\rho_{\star} & = & \frac{M_{\star}}{\frac{4}{3}\pi\,R_{\star}^3}, \\
\end{eqnarray*}

\noindent
we obtain Eq.~1:

\begin{eqnarray*}
z^2(t)\, & = & \Big(\frac{8\,\pi^2\,\rm G}{3\,\rm P}\Big)^{2/3}\,\rho_{\star}^{2/3}\,\Big[(t-T_0)^2-\Big(\frac{T_d}{2}\Big)^2\Big]\,+ \nonumber \\
         & + & \,(1\,+\,r)^2. \nonumber \\
\end{eqnarray*}

\section*{Appendix B: RM effect during the ingress and the egress}

In this  Appendix we  introduce a new  analytic representation  of the
Rossiter-McLaughlin effect valid during the ingress and the egress of
the  transit, that  is between  the first  to the  second  contact and
between  the third  to the  fourth  contact. During  these phases  the
formula presented by Hirano et al.~(2010) accounts for the velocity of
the star  below the disk  of the planet  considering the value  of the
velocity at the center of the  disk of the planet, and it is therefore
valid for the small  planets approximation.  We instead integrated the
velocity profile  below the disk  and calculated the  average velocity
which makes our approach consistent  with the calculation of Hirano et
al.~(2010) for the remaining phases of the transit. This derivation is
based on the method described in (P\'al 2012).

Therefore if we define X and Y as the coordinates of the center of the
planet  at a  given instant  during  the transit,  accordingly to  the
choice of parameters we adopted in this paper we have

\begin{figure}
\begin{center}
\includegraphics[width=\linewidth]{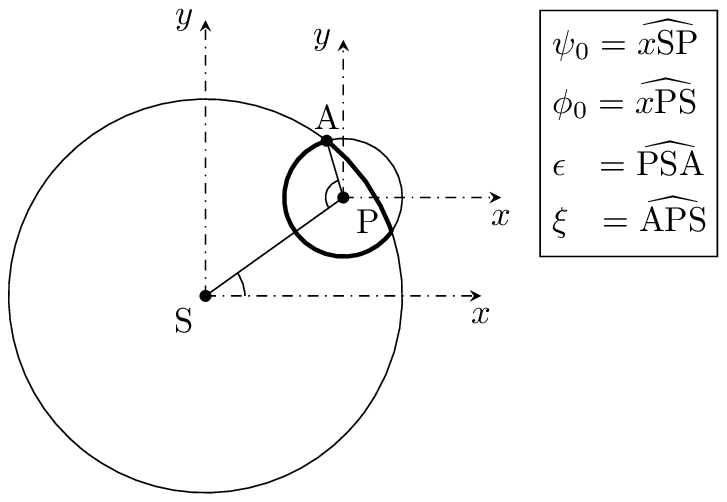}
\caption{\label{fig:coord} 
Definition of the  $\psi_0$,   $\phi_0$,  $\epsilon$  and $\xi$ angles 
introduced in the text.
}
\end{center}
\end{figure}

%
%
\begin{eqnarray*}
X   & = & \frac{(1-e\,\sin \omega)}{\sqrt{1-e^2}}  \Big(\frac{8  \pi^2  G}{3  P}\Big)^{1/3}\,\rho^{1/3}_{\star}\,(t-T_0)\ , \nonumber\\
\nonumber \\
Y   & = & \frac{(1-e\,\sin \omega)}{\sqrt{1-e^2}} \sqrt{(1+r)^2  -  \Big(\frac{8  \pi^2  G}{3  P}\Big)^{2/3}\,\rho^{2/3}_{\star}
          \Big(\frac{T_d}{2}\Big)^2}\ . \nonumber \\
\end{eqnarray*}
%
%
Then if  $\lambda$ is the spin-orbit  angle projected on  the plane of
the  sky, in  the rotated  coordinate  system which  vertical axis  is
aligned with  the projected spin axis  of the star,  the coordinates x
and y are given by
\begin{eqnarray*}
x   & = & X  \cos \lambda   -  Y   \sin \lambda\ ,  \nonumber\\
y   & = & X  \sin \lambda   +  Y  \cos \lambda\ .  \nonumber\\
\end{eqnarray*}
Let  $\psi_0$,   $\phi_0$,  $\epsilon$  and  $\xi$  be   defined  as  in
Fig.~\ref{fig:coordinates} with  respect to the  xy rotated coordinate 
system, then 
\begin{eqnarray*}
\psi_0\, & = & \,\rm atan_2(y, x)\ ,
\nonumber \\
\phi_0\, & = & \, \psi_0\,+\,\pi\ .
\end{eqnarray*}
Assuming that the radius of the star is normalized to unity, one gets
\begin{eqnarray*}
\epsilon\, & = & \,\rm acos\Big(\frac{1-r^2+z^2}{2\,z}\Big)\ , \nonumber \\
\nonumber \\
\xi\, & = & \,\rm acos\Big(\frac{r^2+z^2-1}{2\,r\,z}\Big)\ , \nonumber \\
\end{eqnarray*}
where 
\begin{eqnarray*}
z\, & = & \, \sqrt{x^2+y^2}\ . 
\end{eqnarray*}
Defining now the angles $\psi_a$, $\psi_b$, $\phi_a$ and $\phi_b$ as
\begin{eqnarray*}
\psi_a\, & = & \psi_0\,-\,\epsilon\ ,
\nonumber \\
\psi_b\, & = & \psi_0\,+\,\epsilon\ ,
\nonumber \\
\phi_a\, & = & \phi_0\,-\,\xi\ ,
\nonumber \\
\phi_b\, & = & \phi_0\,+\,\xi\ ,
\end{eqnarray*}
and defining the following functions
\begin{eqnarray*}
A(\alpha,\beta,\gamma)   & = &   \frac{1}{2}  \alpha^2  \beta  \sin \gamma  
                              +  \frac{1}{2}  \alpha  \beta^2  \left(\gamma+\frac{1}{2}  \sin 2\gamma \right) \\
                         & + &   \frac{1}{2}  \beta^3  \left(\gamma+\frac{1}{3}  \sin^3 2\gamma \right)\ ,
\nonumber \\
\nonumber \\
B(\alpha,\beta,\gamma)   & = &      \frac{1}{2}  \alpha  \beta  \sin \gamma   
                                 +  \frac{1}{2}  \beta^2  \left(\gamma+\frac{1}{2}  \sin 2\gamma \right)\ , \\
\end{eqnarray*}
the subplanet velocity $v_p$ is given by
\begin{eqnarray}
\frac{v_p}{v\sin i}\,& = & \,\frac{A_{\rm tot}}{B_{\rm tot}} \nonumber \\
\end{eqnarray}
where
\begin{eqnarray*}
A_{\rm tot}  & = & A(0,1,\psi_b)  -  A(0,1,\psi_a)  +  A(x,r,\phi_b)  -  A(x,r,\phi_a)\ , \nonumber \\
\nonumber \\
B_{\rm tot}  & = & B(0,1,\psi_b)  -  B(0,1,\psi_a)  +  B(x,r,\phi_b)  -  B(x,r,\phi_a)\ . \nonumber \\
\end{eqnarray*}
During the full transit pahse, Eq.~10 reduces to
\begin{eqnarray}
\frac{v_p}{v\sin i}  & = & x\ , \nonumber \\
\end{eqnarray}
which is equal to Eq.~A8 of Hirano et al.~(2010).


\end{document}